\documentclass[twocolumn]{aastex62}
\usepackage{amsmath}	
\usepackage{amssymb}
\usepackage{wasysym}
\usepackage{threeparttable}
\usepackage{multirow}
\turnoffediting
\graphicspath{{./}{figures/}}

\shorttitle{Detecting Solar-like Oscillations with ConvNets}
\shortauthors{Hon et al.}

\begin{document}

\title{DETECTING SOLAR-LIKE OSCILLATIONS IN RED GIANTS WITH DEEP LEARNING}

\correspondingauthor{Marc Hon}
\email{mtyh555@uowmail.edu.au}

\author[0000-0003-2400-6960]{Marc Hon}
\affiliation{School of Physics, The University of New South Wales, Sydney NSW 2052, Australia}

\author{Dennis Stello}
\affiliation{School of Physics, The University of New South Wales, Sydney NSW 2052, Australia}
\affiliation{Sydney Institute for Astronomy (SIfA), School of Physics, University of Sydney, NSW 2006, Australia}
\affiliation{Stellar Astrophysics Centre, Department of Physics and Astronomy, Aarhus University, Ny Munkegade 120, DK-8000 Aarhus C, Denmark}

\author{Joel C. Zinn}
\affiliation{Department of Astronomy, The Ohio State University, Columbus, OH 43210, USA}



\begin{abstract}

Time-resolved photometry of tens of thousands of red giant stars from space missions like \textit{Kepler} and K2 has created the need for automated asteroseismic analysis methods.  The first and most fundamental step in such analysis, is to identify which stars show oscillations.  It is critical that this step can be performed with no, or little, detection bias, particularly when performing subsequent ensemble analyses that aim to compare properties of observed stellar populations with those from galactic models.  Yet, an efficient, automated solution to this initial detection step has still not been found, meaning that expert visual inspection of data from each star is required to obtain the highest level of detections.
Hence, to mimic how an expert eye analyses the data, we use supervised deep learning to not only detect oscillations in red giants, but also predict the location of the frequency at maximum power, $\nu_{\mathrm{max}}$, by observing features in 2D images of power spectra. By training on \textit{Kepler} data, we benchmark our deep learning classifier against K2 data that are given detections by the expert eye, achieving a detection accuracy of 98\% on K2 Campaign 6 stars and a detection accuracy of 99\% on K2 Campaign 3 stars. We further find that the estimated uncertainty of our deep learning-based $\nu_{\mathrm{max}}$ predictions is about 5\%. This is comparable to human-level performance using visual inspection. When examining outliers we find that the deep learning results are more likely to provide robust $\nu_{\mathrm{max}}$ estimates than the classical model-fitting method.

\end{abstract}



\section{Introduction}
Since the launch of CoRoT \citep{Baglin_2006} and \textit{Kepler} \citep{Borucki2010}, the number of stars that can be probed by asteroseismology has been increasing dramatically, particularly for red giants.  As a result, the study of solar-like oscillations in giants has evolved into a field dominated by ensemble analyses \citep{Chaplin_2013,Hekker_2017,Garcia_2018}, which has opened up prospects to study large stellar populations in great detail to inform galactic archaeology studies \citep{Miglio_2012, Casagrande_2014, Aguirre_2017}. Of particular importance is \textit{Kepler}'s second-life mission, K2 \citep{Howell_2014}, which provides high-precision photometry of tens of thousands of red giants along the ecliptic.  While light curves from K2 only span around 80 days, the mission's 360-degree ecliptic coverage proves highly valuable for galactic archaeology.  In particular, asteroseismology in tandem with ground-based measurements of stellar surface properties (e.g \citealt{Martell_2016,Traven_2017, Wittenmyer_2017}) of K2 targets allows the inference of fundamental stellar parameters such as the radius, mass, distance, and age for stars across the Galaxy never before probed by asteroseismic means.  Driven by this great opportunity, the K2 Galactic Archaeology Program (GAP) \citep{Stello_2015} specifically targets 4,000-10,000 potentially oscillating giants from each K2 campaign \citep{Stello_2017}.
Hence, it has become critical that the analysis of these data can be performed more or less automatically, because manual analysis and expert visual inspection becomes far too time consuming and introduces subjective decision making which limits `reproducibility' of the analysis.   

Fortunately, the large amount of high-quality data, particularly from \textit{Kepler}, has sparked developments of increasingly more sophisticated analysis methods, some fully automated (e.g. \citealt{SYD,Mosser_2009,Hekker_2010,Mathur_2010,Kallinger_2016}). These methods typically use high-dimensional parametric model fitting, and some are based on state-of-the-art Bayesian fitting coupled with Markov chain Monte Carlo techniques to obtain statistically rigorous results including robust uncertainties. They are now capable of measuring a range of seismic properties of stars, including the frequency of maximum power, $\nu_\mathrm{max}$ (which gives strong constraints on surface gravity, $\log g$), and the overtone frequency spacing of acoustic modes, $\Delta\nu$ (which estimates stellar mean density). 


However, the most fundamental result, assessing whether or not a given star shows oscillations in the data, has turned out to be one of the hardest problems to solve.  The automated analysis pipelines use different levels of statistical sophistication to assess whether oscillations are detected, but in general they focus more on providing accurate seismic measurements of stars that are `clearly' oscillating, rather than giving robust and unbiased assessments of which stars do, or do not, oscillate.  

Comparisons across pipelines illustrate clearly that they have different detection biases when applied to K2 data \citep{Stello_2017}.  Interestingly, when benchmarked against expert visual inspection of the data, two results emerge: (1) not a single pipeline provides as many detections as the `expert eye' and (2) even the combined result from multiple pipelines does not reach the level obtained by the expert's visual assessment. 
The resulting poorly defined detection biases from automated pipelines affect galactic archaeology studies particularly badly.  We are therefore in a situation where comparisons between observed and galactic model-synthesised stellar property distributions are currently limited by our inability to automatically and robustly assess if a given star shows oscillations.  This detection bias problem  has recently been highlighted by Zinn et al. (in prep), who showed a much better match of a galactic model-based stellar population with the expert's visual-based results than with the result from the automated pipelines.

It seems that the natural variation in real-world noise and signal makes the difference between detection and non-detection too blurry when relying fully on prescribed parametric fits to the data, because we lack a comprehensive mathematical description of the data. The expert's brain, however, has more free parameters and can therefore more easily learn to distinguish detection and non-detection.  Hence, to mitigate the detection bias shortcoming of current automated pipelines, we set out to develop an automated method that can efficiently mimic the expert's visual approach.

Here, we turn to the use of machine learning to perform such a task due to its high efficiency in performing automated data analyses. In asteroseismology for instance, machine learning has proven effective in numerous applications, such as the estimation of $\log g$ for giant stars \citep{Liu_2015} and the prediction of fundamental parameters of main sequence stars \citep{Bellinger_2016}. However, to specifically mimic the expert's visual approach, we use deep learning, which is a sub-category of machine learning methods that acts as an approach to artificial intelligence. Compared to conventional machine learning methods which learn highly task-specific properties from data, deep learning learns features and patterns from data, such that it is capable of `understanding' abstract data representations. While generally still in its infancy in astrophysics, it has recently shown great promise in some applications, such as classifying the evolutionary stage of red giants \citep{Hon_2017,Hon_2018}, star-galaxy classification \citep{Kim_2016}, and even detecting exoplanet transit signals \citep{Shallue_2017}. 

In this paper, we develop a deep learning model that detects the presence of red giant oscillations, and another that provides an estimate of the frequency at maximum oscillation power, $\nu_{\mathrm{max}}$. We represent the frequency power spectrum in the form of a 2D image, such that the detection of red giant oscillations can be formulated as an image recognition task, which has seen much success with the use of deep learning (e.g \citealt{Kriz_2012, Szegedy_2015}). The estimation of $\nu_{\mathrm{max}}$, however, can be attributed to the task of object localization, where the coordinates of the object of interest on an image need to be predicted (e.g \citealt{Szegedy_2013, Liu_2016}). With the use of deep learning, object localization too has achieved a high level of performance (e.g. \citealt{Sermanet_2014}). Hence, in this paper, we show that this novel approach of analysing 2D images of power spectra with deep learning can be a highly viable method in asteroseismic detections.

\section{Methods}
We first introduce the data we use for training and testing, followed by a description of how the data is prepared as input for our deep learning models. Here, we use the term `model' to refer to a fully-trained neural network. We then introduce the concept of deep learning, and detail the structure and parameters of our deep learning models, namely the classifier model for detecting solar-like oscillations in power spectra and the regression model for estimating $\nu_{\mathrm{max}}$

\subsection{Data}
\label{sec:data}
\begin{figure}
	\centering
	\includegraphics[width=\linewidth]{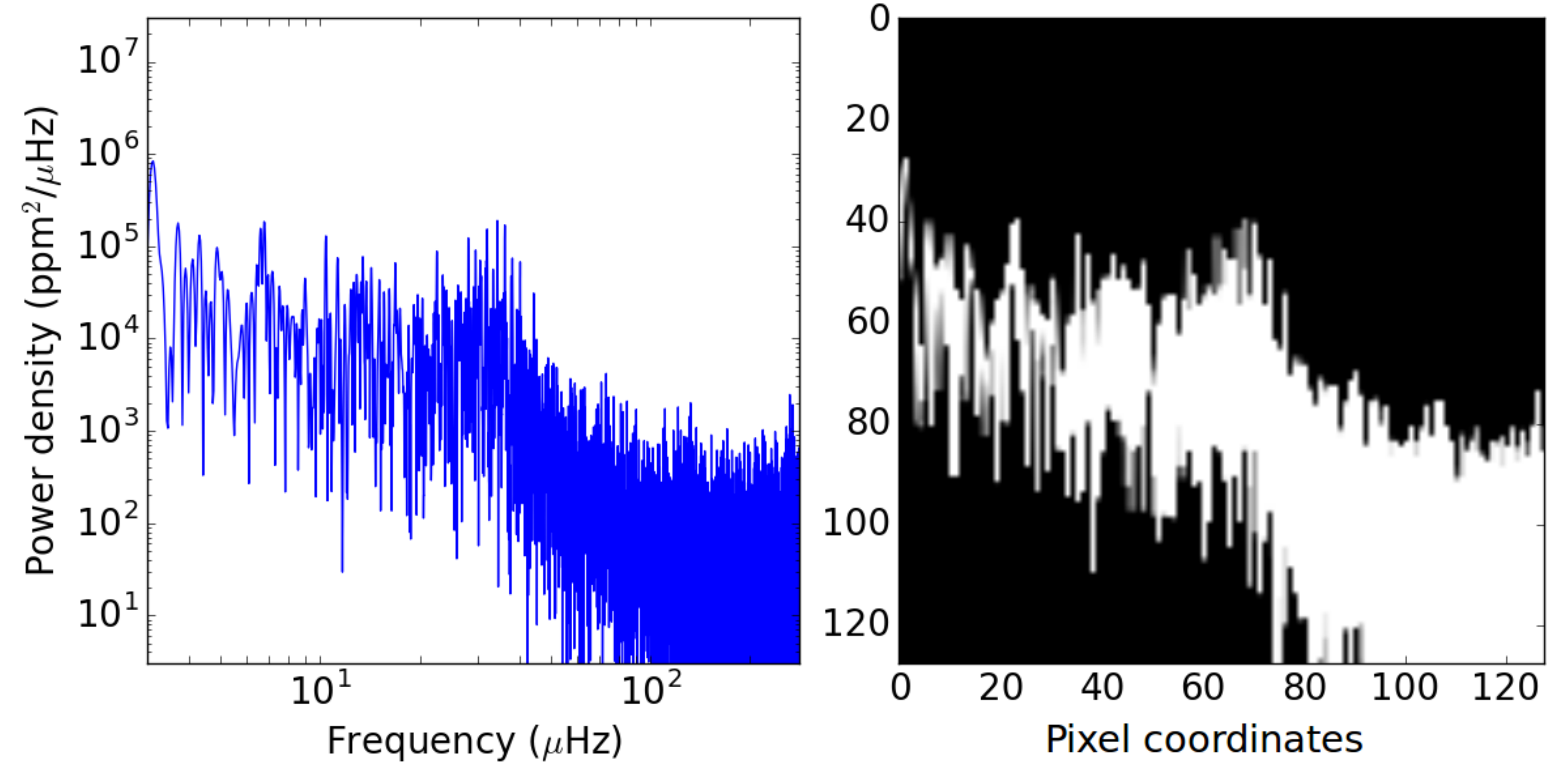}
	\caption{(Left) Power density spectrum of \textit{Kepler} red giant KIC 12072067 and its corresponding 128 X 128 pixel grayscale 2D image (right). The axes labels for the 2D image are pixel coordinate values.}
	\label{processimage}
\end{figure}

\subsubsection{Training Set}
To train our classifier model, we use a training set comprising 31123 \textit{Kepler} targets from end-of-mission long-cadence data ($\Delta T \simeq 30\,$min). Of these, 15924 are known oscillating red giants from the analysis by \citet{Yu_2018}, while the remaining 15199 stars do not show solar-like oscillations in the frequency range $\nu \apprle 283\mu$Hz, which we denote as non-detections. We obtain these non-detections by randomly selecting, in Kepler Input Catalog (KIC) ID, a subset of stars from the \textit{Kepler} Mission Data Release 25 catalog \citep{Mathur_2017} that have input log $g$ values not taken from asteroseismology. We have further verified by visual inspection of their power spectra that they do not show detectable red giant oscillations. Because the current application of our deep learning models will be focused on K2 data, we use 82-day segments from the \textit{Kepler} time series to construct the power spectra so they approximate K2 data. 

For our regression model, we only use the 15924 \textit{Kepler} oscillating red giants, with their $\nu_{\mathrm{max}}$ values taken from the red giant catalogue by \citet{Yu_2018}, which were derived using the SYD pipeline \citep{SYD}. To improve the prediction performance of the regression model at high and low $\nu_{\mathrm{max}}$ where our red giant sample is underrepresented, we artificially boost the number of training stars with $\nu_{\mathrm{max}} < 5.5\mu$Hz and $220\mu$Hz$<\nu_{\mathrm{max}} < 283\mu$Hz by using multiple independent 82-day segments from each star in those ranges. As discussed by \citet{Hon_2018}, the power density spectra of different segments of the same star appear similar to one another but still show natural variations in noise. In the end, we have a total of 19764 time series to train the regression model.

\subsubsection{Validation Set} 

To select the best classifier model that generalize well to unseen data, we adjust the model parameters (such as the number of network layers, see Section \ref{classmodel} for details) and measure its resulting performance on a validation set during training. To ensure that our model can predict well across all K2 campaigns, we wish to benchmark its performance on a validation set that is representative of K2 data. Hence, we use a validation set comprising 7196 stars from Campaign 6 of the K2 mission that are given classifications by visual inspection. 
The inspection was performed on plots of power density spectra similar to Figure 1 (left) by an expert as described in Sect 2.3 (see also \citealt{Stello_2017}; their Sect 4.1). The spectra were based on EVEREST light curves \citep{Luger_2016}. We inspected all 8312 stars from the K2 Galactic Archaeology Program target list \citep{Stello_2015,Stello_2017}.
From this set of stars, 2743 show red giant oscillations, while 4453 do not. For the remaining stars, the inspection delivered an ambiguous detection label, and they were therefore discarded in the following.  

Out of the 2743 oscillating red giants, 1747 have measured $\nu_{\mathrm{max}}$ values from the BAM pipeline (See summary in Section 3.1 of \citealt{Stello_2017}), which we use as the validation set for our regression model.

\subsubsection{Test Set}
 
As a final, unbiased measure of the classifier performance, we measure the classifier performance (which has been tuned for optimal performance on the validation set) on a test set. Our test set consists of 12865 stars from K2 Campaign 3. 
This set originates from visual inspection of all 14151 stars for which there are long cadence EVEREST light curves on MAST\footnote{\url{https://archive.stsci.edu/}}. They include all K2 Galactic Archaeology Program targets.  Here we used the consensus detection classification by two independent experts, which resulted in 1370 positive detections, and 11495 non-detections.  
Out of the 1370 positive detections, 525 have $\nu_{\mathrm{max}}$ values measured by the BAM pipeline, which we use as the test stars for our regression model. 

\edit1{While we test and validate only on K2 data, we expect our test results to be relevant for TESS data as well. This is because data from TESS are expected to have similar photometric properties to \textit{Kepler} data, but for stars about 5 magnitudes brighter \citep{TESS}, while K2 data in turn has a comparable level of photometric precision to that of \textit{Kepler} data for stars of the same brightness \citep{Luger_2016}. Thus, we expect our model performances on the K2-based test set to be representative of their performances on 82-day TESS data as well. However, a caveat with our method of using simulated data for training is that any TESS-specific instrumental artifacts in the data are not considered. However, these can only be taken into account once the first full year of TESS data is obtained, for which we can then take the full length 1-year time series, which will provide the truth labels, and then sub-divide them into shorter lengths for training and testing.}

\subsection{Data Preparation}

To maximize the efficiency of the deep learning classifier in detecting the oscillations, we attempt to reduce the effect from the spectral window on the power spectrum, which can otherwise cause high-power at low-frequency to leak into higher frequency bins. In effect, the leakage obscures/washes out the intrinsic global shape of the frequency distribution of granulation, oscillations, and white noise \citep{Garcia_2014b}.  To reduce this effect we use two methods. First, we apply a high-pass boxcar filter with a width of 2 days similar to that done by \citet{Stello_2015}, which results in an approximate high-pass cut-off frequency of $3\mu$Hz, below which we can not resolve the oscillations from K2 data anyway. 
Next, we fill the gaps in the time series 
using the inpainting technique described by \citet{Pires_2015}. 

The next step involves the conversion of the data into a 2D input image for the deep learning models. We plot the power density spectrum of each star with logarithmic axes within the ranges $3\mu$Hz $\leq \nu \leq283\mu$Hz and 3 ppm$^2\mu$Hz$^{-1}<$ $P$ $<3\times 10^7$ppm$^2\mu$Hz$^{-1}$, where $P$ is power density. This power density range is sufficient to capture the range of oscillation and granulation power within the listed frequency range (see \citealt{Mathur_2011}, their Figure 3). The 2D input image is a grayscale snapshot of this plot plotted in white against a black background. Figure \ref{processimage} shows an example of the power density spectrum along with its corresponding 2D image. Before using the image as input to the deep learning network, we pixelate/tile it into 128 equal-size pixels on a side with values ranging from 0 (black) to 255 (white).

\begin{figure}
	\centering
	\includegraphics[width=\linewidth]{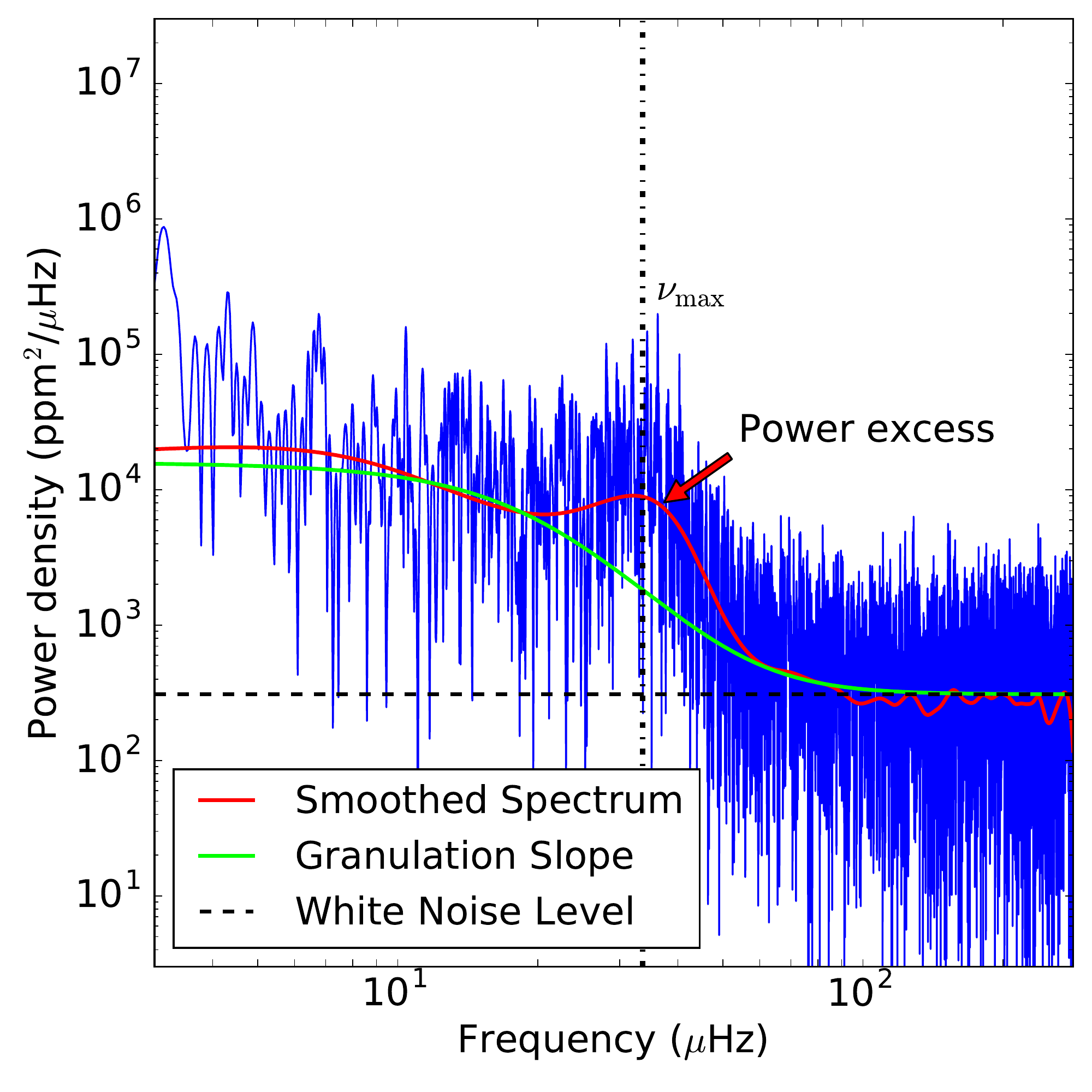}
	\caption{Log-log power density spectrum of KIC 12072067 showing a combination of a sloping granulation profile (green), a flat white noise component (dotted black line), and a power excess envelope containing solar-like oscillation modes. The frequency at maximum oscillation power, $\nu_{\mathrm{max}}$ (dashed dotted line), is the estimated center of the power excess. The red curve is a heavily smoothed version of the spectrum to guide the eye.}
	\label{granulation}
\end{figure}

\begin{figure*}
	\centering
	\includegraphics[width=0.9\linewidth]{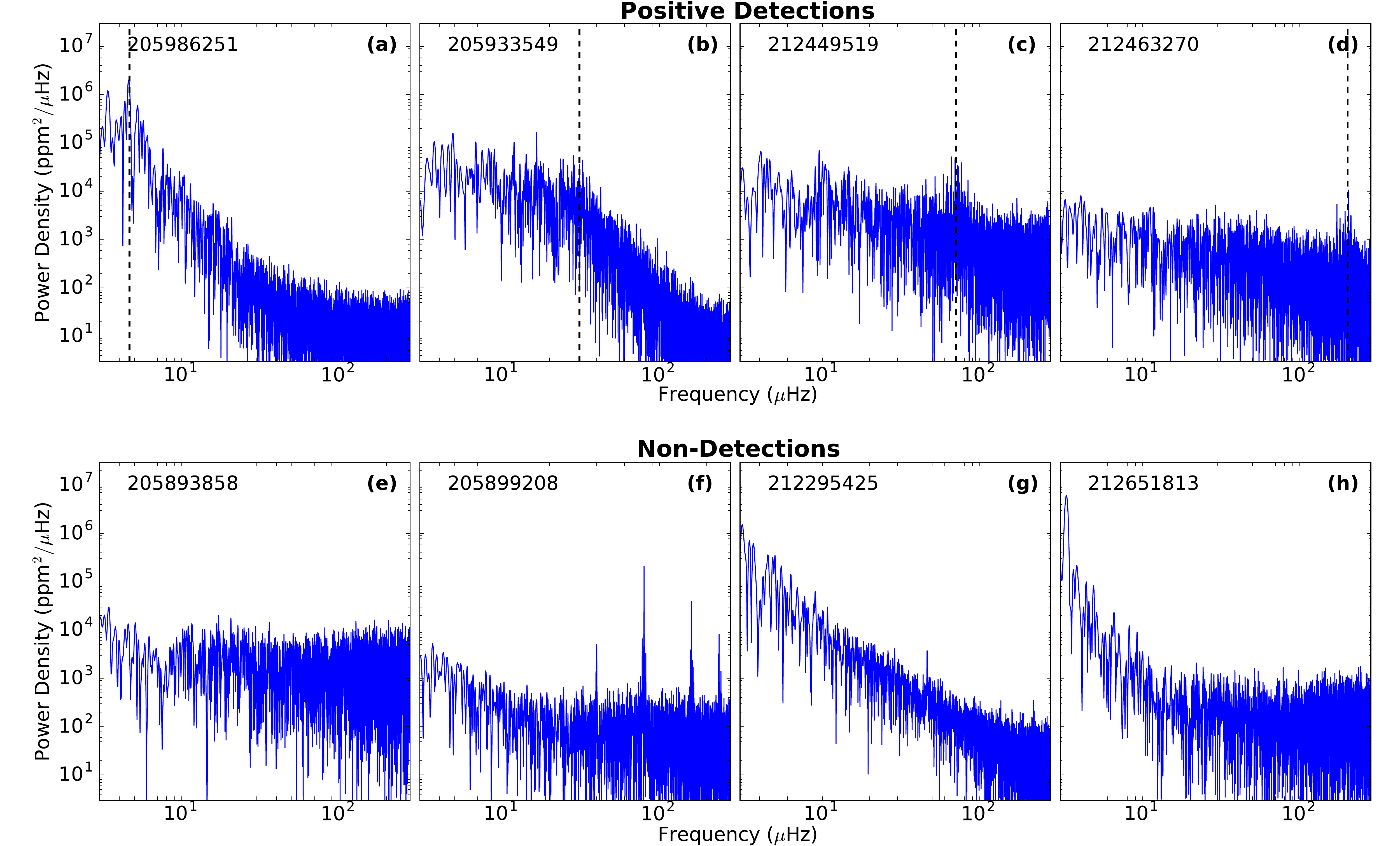}
	\caption{Examples of power density spectra of stars observed by K2 from Campaigns 3 and 6 showing solar-like oscillations (top row) and those that do not (bottom row) according to visual inspection. Each image is labelled with the K2 Ecliptic Plane Input Catalog (EPIC) ID \citep{Huber_2016} of the star. The classification criteria is as follows: (a-d) Each power spectra profile shows a clear granulation slope with a visible power excess. The dashed vertical lines are visually-determined estimates of $\nu_{\mathrm{max}}$ to guide the eye. (e) The profile shows a high white noise level, with no features indicating the presence of oscillations or granulation. (f) No granulation slope or `Gaussian-like' power excess, however other forms of variability are present. (g) Presence of spectral leakage due to high-power oscillations at frequencies below the plotted range. No power excess is visible. (h) Presence of a steep granulation slope at low frequencies, indicating potential $\nu_{\mathrm{max}}$ at very low frequencies. However, an oscillation power excess cannot be seen within the plot's frequency range.}
	\label{comparisons}
\end{figure*}

\subsection{Image Features}
Here we describe criteria in the form of visual characteristics or `features' within the power spectra that we use to visually determine the presence of solar-like oscillations. A red giant that clearly shows solar-like oscillations (a positive detection) will have a power spectrum profile consisting of a `Gaussian-like' excess hump of power on top of a downward-sloping granulation background, along with the possible addition of a flat white noise level. An example of this profile is shown in Figure \ref{granulation}.

When manually assigning the ground truth labels, we require the clear presence of both the granulation background and the oscillation power excess hump to convincingly classify a power spectrum as a positive detection. We also observe the smoothed version of the spectrum to aid the detection of the power excess by eye. Other contextual clues for visual classification include the links between oscillation and granulation time scales and amplitudes (e.g. \citealt{Huber_2011,Kjeldsen_2011, Mathur_2011, Yu_2018}). This implies that the relative position of the power excess within the 2D image also plays an important role. For example, we expect lower $\nu_{\mathrm{max}}$ stars to have higher-amplitude and lower-frequency granulation profiles, with their oscillation power excesses occupying a higher position in the 2D image. There are cases in which a high-amplitude low-frequency granulation profile is evident without the  presence of oscillation power excess. While these are likely to be stars with $\nu_{\mathrm{max}}<3\mu$Hz, we nonetheless categorize these as non-detections. We show typical examples of power spectra classified as positive detections and non-detections in Figure \ref{comparisons}.  

Finally, we note that the visual inspection process could introduce some incorrect truth labels due to the tedious and extremely repetitive task involved on such large data sets and the speed by which it had to be performed to be manageable. However, our previous experience with deep learning architectures similar to what we use here suggest the trained classifier's performance will not heavily suffer from a moderate amount of incorrect truth labels \citep{Hon_2018}. 


The deep learning models that we develop in this study are therefore expected to perform their required tasks by learning morphological features from the 2D images such as the shape of a power excess. For instance, the deep learning models should recognize the presence of granulation by identifying the shape and position of the downward sloping profile within the image.  Because our deep learning models do not require any form of model fitting on power spectra, our method is robust such that it is capable of predicting on power spectra showing different profiles (morphologies). This ability to learn different representations of power spectra enables it to better account for the full realistic variations in the data as compared to fitting a mathematical model with many fewer free parameters.

\subsection{Deep Learning Models}
\begin{figure}
	\centering
	\includegraphics[width=\linewidth]{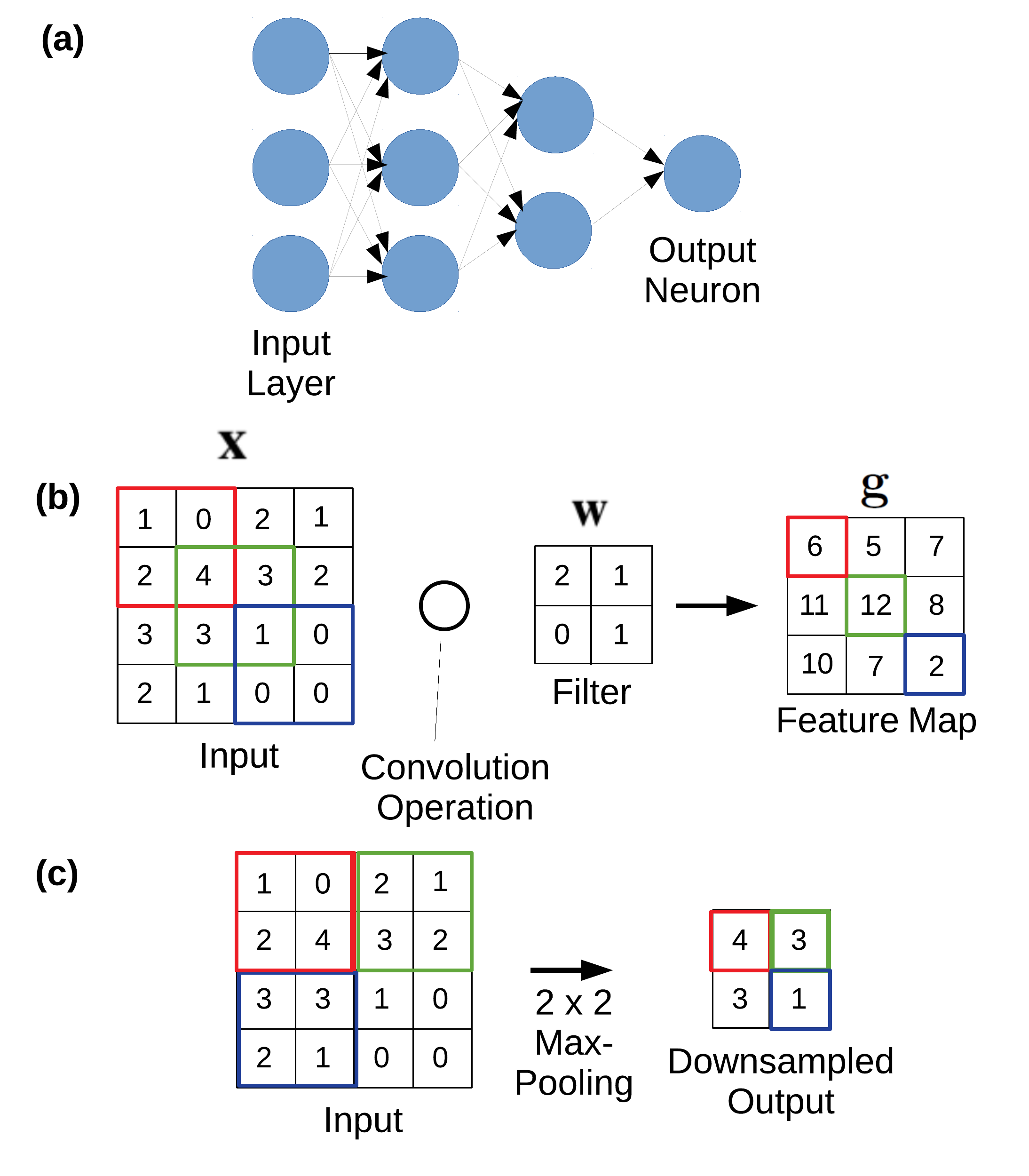}
	\caption{(a) A fully-connected deep neural network with 3 input neurons and 1 output neuron. The weights connecting each neuron in the network are represented by arrows. (b) The convolution operation within a 2D convolutional layer. The tile-like structure of the input, $\mathrm{\textbf{x}}$, is analogous to a collection of neurons in a layer, and the layer weights $\mathrm{\textbf{w}}$ act as a filter. When the values in color-highlighted tiles in the input are convolved with the filter, they produce values in tiles of the feature map highlighted by the same color. (c) Similar to (b), except that the operation performed is 2 x 2 max-pooling with no overlap between pooled regions in the input.}
	\label{convdiagram}
\end{figure}

Deep learning is typically applied using deep neural networks, which comprise a collection of artificial neurons that are connected to one another, simulating a biological neural network. These artificial neurons contain real numbers and signal to one another by connections in the form of real-numbered weights. A common way of structuring a deep neural network is to have multiple levels or layers of neurons, where neurons in each layer are connected to those in the next layer to form a network of stacked layers, forming a \textit{fully-connected} network (Figure \ref{convdiagram}a). Except for the first layer, the total input to a neuron in every network layer in a deep neural network is computed by a linear combination of weights and values from connected neurons from the previous layer. The output of each artificial neuron is then obtained by computing a non-linear function of its total input, known as an \textit{activation function} \citep{Blatt62}. Along with the multiple connections between many neurons in a deep neural network, the computed non-linearities allow the neural network to approximate complex relations between the network input and output. The output of the network is hence computed by passing the network inputs forward through layers of the network. 

In this study, we use a convolutional neural network \cite{LeCun1998}, which is a variation of the deep neural network that specializes in the extraction of features from data with a grid-like topology such as 2D images. Feature extraction is done through a \textit{convolutional} layer, where the weights, $\mathrm{\textbf{w}}$, that normally connect neurons in adjacent layers are now made to act as a convolutional filter. This is done by constraining a combination of weights to be shared among all the neurons in a layer.  By sliding the filter across neurons in the convolutional layer, $\mathrm{\textbf{x}}$, a kernel convolution is effectively performed, which extracts a particular feature from the data depending on the values of the combination of weights within the convolutional filter (see Figure \ref{convdiagram}b). The output of the convolution across the layer is then stored in a \textit{feature map} \citep{Rumelhart1988a}, $\mathrm{\textbf{g}}$, where each contains the result of the extraction of a specific feature. To be able to extract many different features, a convolutional layer uses multiple filters, implying  multiple combinations of weights for the layer. \edit1{In particular, the output in the $l$-th feature map from the $l$-th convolutional filter $\boldmath{\mathrm{w}^{(l)}}$ is given by:}

\begin{equation}
g^{(l)} = f\bigg(\sum_{i=0}^{m}\boldmath{\mathrm{w}}^{(l)} \circ \ \mathrm{x}_{i}\bigg),
\label{convEq}
\end{equation}
\edit1{with $m$ denoting the number of stars in the data set, and $f$ denoting the activation function.}

\begin{figure*}
	\centering
	\includegraphics[width=\linewidth]{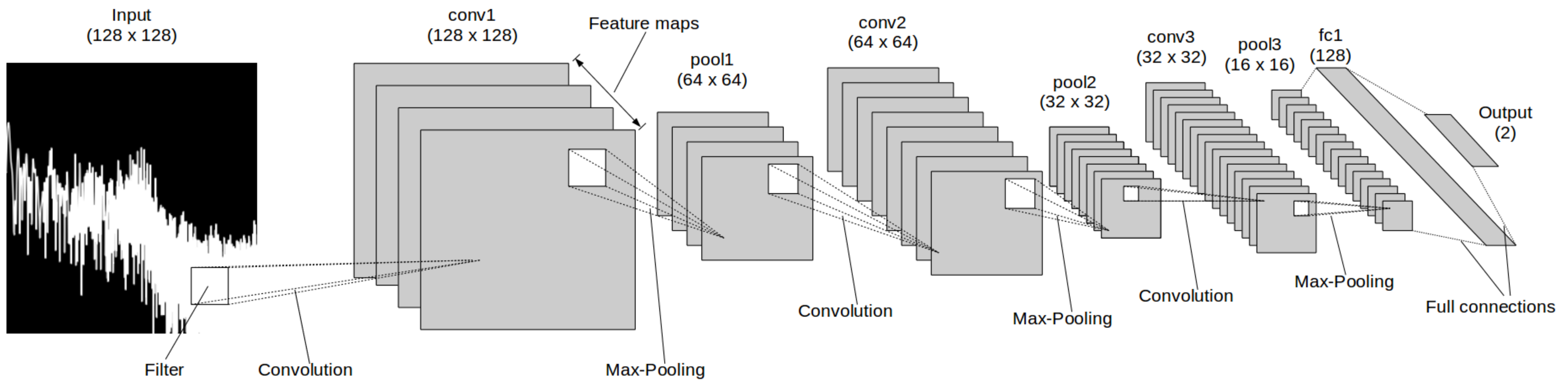}
	\caption{Schematic for the 2D convolutional neural network classifier. For convolutional (conv) and pooling (pool) layers, the values in brackets indicate the dimension of each feature map, while for fully-connected (fc) layers, they indicate the number of neurons within that layer.}
	\label{convschematic}
\end{figure*}

Following a convolutional layer, a \textit{pooling} layer is commonly applied as a way of downsampling the output of the convolutional layer. We use max-pooling to downsample, which only retains the neuron with the maximum value within groups of adjacent neurons within the layer. An example of this is shown in Figure \ref{convdiagram}c. In principle, pooling reduces the complexity of the neural network because the fewer the number of resulting neurons, the fewer the number of weights and computational steps required. It also helps the network to achieve spatial invariance to the position of features within the data \citep{Bengio2013}. From convolutional and pooling layers, a convolutional neural network can then be constructed by alternating these two layers , as can be seen in Figure \ref{convschematic}. Such a configuration generally allows the network to be able to extract multiple highly abstract features from data.

We use 2D convolutional neural networks in our study, which perform 2D convolutions to learn 2D filters and extract features from 2D images of the power spectra. However, in order for the neural network to learn how to perform a specific task, it has to be trained on data. In this study, we train our neural networks by \textit{supervised learning}, which is an iterative procedure where the network is provided with ground truth values and learns how to predict correctly by adjusting its outputs to match the ground truth. In neural networks, this adjustment is performed by gradient descent, where the network weights are iteratively updated \edit1{using} the derivative of the output error with respect to the magnitude of the weights. From the output layer, this error derivative is propagated backwards throughout the network by the chain rule of calculus \citep{Rumelhart1986}, such that all network weights are updated for every training iteration. In the context of a convolutional neural network, during training the network effectively learns specific filters that extracts features from the input data that allow it to perform accurate prediction tasks. For our classifier model, this task is predicting classification labels, while for our regression model, this task is predicting a real number. In the following sections, we describe in detail the structure and network parameters of these two models.  

\subsubsection{Classifier Model}\label{classmodel}
	\begin{table} 
		\centering
		\caption{A sequential list of network layers and parameters in the classifier model. The Size column refers to the size of the filter at each layer.
		}
		\label{convtable}
        \begin{tabular}{|c|c|c|c|c|}
        \hline
        Layer & Feature maps & Size & Activation&Dimension\\
        \hline
        Input & - & - & -&128x128\\
        drop1 & - & - &-&128x128\\
        conv1 & 4 & 7x7 & LReLU&128x128\\
        pool1 & 4 & 2x2& -&64x64\\
        conv2 & 8 & 5x5 & LReLU&64x64\\
        pool2 & 8 & 2x2 & -&32x32\\
        conv3 & 16 & 3x3 & LReLU&32x32\\
        pool3 & 16 & 2x2 & -&16x16\\
        drop2 & - & - & -&16x16\\
        fc1 & 1 & - & ReLU&128\\
        Output & 1 & - & Softmax&2\\
        \hline
        \end{tabular}
	\end{table} 
We show the general schematic of the 2D convolutional neural network for the classifier model in Figure \ref{convschematic}. The network consists of 3 convolutional (conv) and pooling (pool) layers stacked sequentially, followed by a fully-connected (fc) layer separating the final pooling layer from the final output layer. In addition to the layers in Figure \ref{convschematic}, we also apply dropout \citep{Sriva2014}, which is a way of preventing overfitting by setting the values of neurons in a layer to zero with a probability $p_\mathrm{drop}$. We use dropout with $p_\mathrm{drop}=0.25$ after the input (drop1) and with $p_\mathrm{drop}=0.5$ (drop2) after the final pooling layer. We additionally use L2 weight penalties with a coefficient of $\lambda=7.5\times10^{-4}$ at each convolutional and fully-connected layer. This form of weight penalty adds the sum of a network layer's squared weights to the computed error in that layer during training, which effectively reduces the magnitude of layer weights and reduces overfitting. Details of each layer such as filter sizes and layer activations are provided in Table \ref{convtable}. For intermediate fully connected layers, we use the Rectified Linear Unit (ReLU) activation function \citep{Nair_2010} defined by $f(x)=$max(0,$x$), where $x$ is the layer input. For convolutional layers, we instead use the Leaky Rectified Linear Unit (LReLU) activation function \citep{Maas_2013}, defined as \begin{equation}
f(x) = 
	\begin{cases}
	x & \text{for } x \geq 0\\
	0.1x & \text{for } x < 0.
	\end{cases}
\end{equation}
Finally, at the output layer, we use the softmax activation function, which gives the probability for the occurrence of each class $j$, where $j$ is either a non-detection (class 0) or a positive detection (class 1). Thus, the output is given by the following:

\begin{equation}
p(y=j|\mathrm{\textbf{x}}) =\frac{e^{\mathrm{\textbf{x}}\cdot\mathrm{\textbf{w}}_j}}{\sum_{k=1}^{2}e^{\mathrm{\textbf{x}}\cdot\mathrm{\textbf{w}}_k}},
\label{softmax}
\end{equation}
where $\mathrm{\textbf{x}}$ are input values to the output layer and $\mathrm{\textbf{w}}$ are the weights of the output layer. Intermediate $p$ values ($p\simeq0.5$) indicate that the classifier is not confident in determining whether the power spectrum  clearly shows a non-detection or a detection of solar-like oscillations, while $p$ values close to 0 or 1 indicate a high confidence in classifying a non-detection or a detection, respectively. 

The objective function to minimize when training the classifier is the cross-entropy or log loss \citep{Murphy_2012}, $E$, given as 
\begin{equation}\label{logloss}
E(\mathbf{y, \hat{y}}) = -\frac{1}{m} \sum_{i=1}^{m} \bigg[y_i\log\hat{y}_i + (1-y_i)\log(1-\hat{y}_i)\bigg],
\end{equation}
where $y$ is the ground truth label, $\hat{y}$ is the predicted probability, and $m$ is the number of stars in the training set. Log loss has a minimum (optimal) value of zero and increases the more incorrect classifications there are in the set of stars. The increase is larger for probabilities that show greater differences from the ground truth value. Hence, we can maximize the accuracy of our classifier by minimizing the log loss during training.  We train three identical classifiers, each with different parameter initializations. Our classification output is the average prediction between these classifier models. This is known as `model averaging', and it further reduces overfitting the data because each model initialization can result in different optimization paths and different ways of forming pattern generalizations when learning \citep{Hansen_1990}. Thus, this provides a more robust classification than the output of a single model.

\begin{figure}
	\centering
	\includegraphics[width=\linewidth]{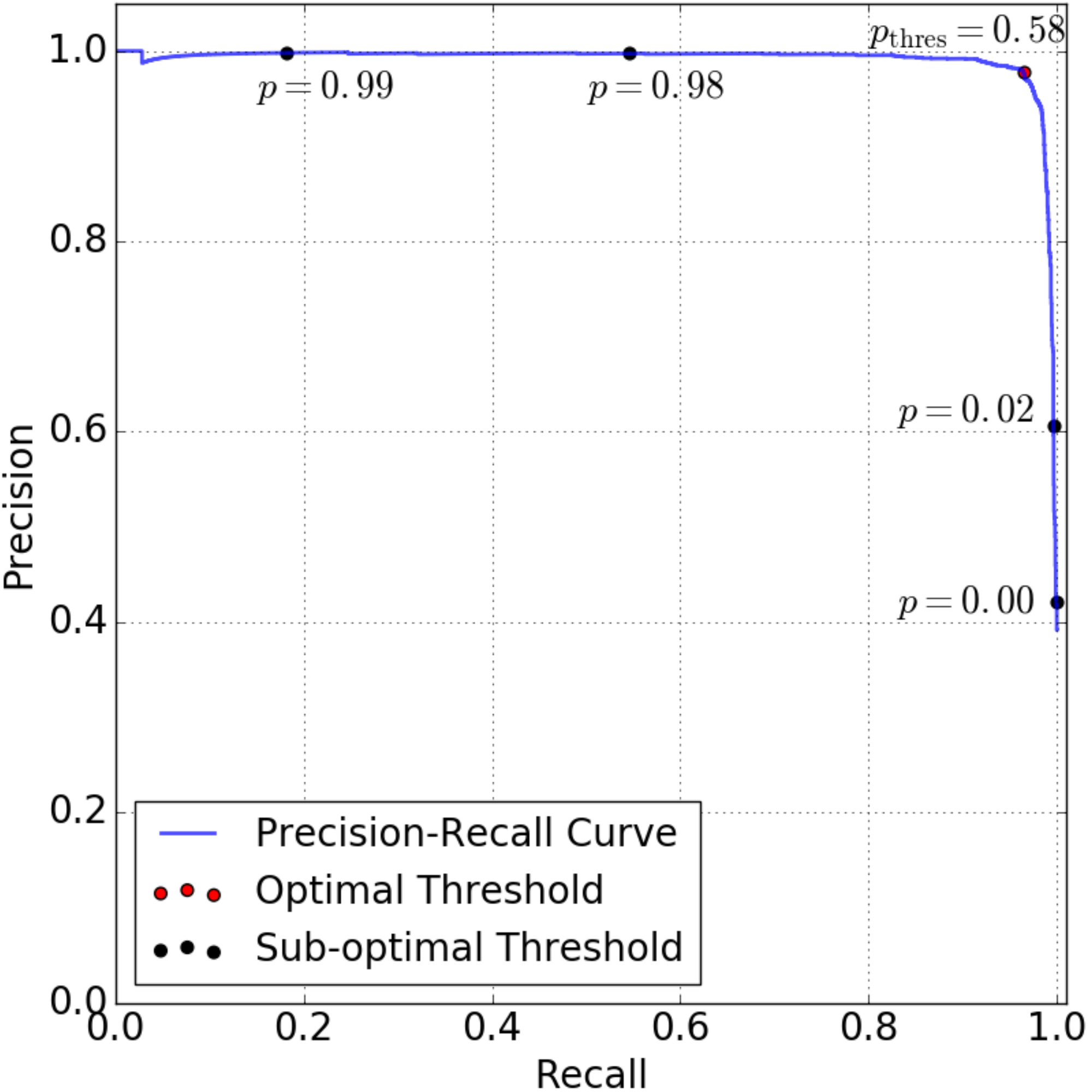}
	\caption{Precision-recall curve of the validation set. The probability threshold that maximizes the sum of precision and recall values is plotted in red. A few other probability thresholds in black are plotted to demonstrate the order of plotting. In general, too high or too low a threshold results in a suboptimal recall or precision.}
	\label{precision-recall-curve}
\end{figure}

By performing a manual search across more than 100 combinations of model structure and parameters, we choose the configuration and parameter values in Table \ref{convtable} by selecting the combination showing the best performance (smallest log loss) on the validation set, with $E \simeq 0.075$.
Besides determining the model structure and parameters, we also use the validation set to determine the optimal probability threshold for separating detections and non-detections. To do this, we construct a \textit{precision-recall curve}. The precision is the ratio of all correctly predicted detections to all predicted detections, while the recall is the ratio of all correctly predicted detections to all true detections. A precision-recall curve shows the precision and recall of the classifier at each probability threshold. We plot this curve as shown in Figure \ref{precision-recall-curve} and select the optimal threshold as the probability threshold that optimizes the sum of precision and recall values, which we find to be $p_{\mathrm{thres}}=0.58$. 

During test time, our final prediction values are the mean of 10 forward passes through the `averaged model' with dropout enabled. Hence, neurons connected to dropout layers (drop1/drop2) in the network will be randomly dropped during each forward pass. This is known as Monte Carlo dropout, and it is effectively a Monte Carlo integration over a Gaussian process posterior approximation \citep{Gal_2016}. To obtain the final prediction uncertainty, we calculate the variance for each of the three classifiers across the 10 forward passes, then take the root of their sum.

\subsubsection{Regression Model}
\begin{table} 
	\centering
	\caption{A sequential list of network layers and parameters in the regression model. The Size column refers to the size of the filter at each layer.
	}
	\label{regressiontable}
    \begin{tabular}{|c|c|c|c|c|}
    \hline
    Layer & Feature maps & Size & Activation&Dimension\\
    \hline
    Input & - & - & -&128x128\\
    drop1 & - & - &-&128x128\\
    conv1 & 4 & 5x5 & LReLU&128x128\\
    pool1 & 4 & 2x2& -&64x64\\
    conv2 & 8 & 3x3 & LReLU&64x64\\
    pool2 & 8 & 2x2 & -&32x32\\
    conv3 & 16 & 2x2 & LReLU&32x32\\
    pool3 & 16 & 2x2 & -&16x16\\
    drop2 & - & - & -&16x16\\
    fc1 & 1 & - & ReLU&1024\\
    fc2 & 1 & - & ReLU&128\\
    Output & 1 & - & Linear&1\\
    \hline
    \end{tabular}
\end{table} 

After determining which stars oscillate, we pass them to our regression model, which provides an estimate of $\nu_{\mathrm{max}}$ from the 2D images of the power spectra.  Rather than using $\nu_{\mathrm{max}}$ values directly, we use pixel coordinates, such that pixel x-coordinates of 0 and 128 correspond to $\nu=3\mu$Hz and $\nu=283\mu$Hz, respectively. In principle, this should simplify the task because the regression model now predicts values on a linear scale instead of a logarithmic scale. Hence, the regression model identifies the position of the power excess within the 2D image and predicts its pixel x-coordinate, $x$, which we then convert into $\nu_{\mathrm{max}}$ using the following conversion formula:
\begin{equation}
\nu_{\mathrm{max}}(\mu\mathrm{Hz})=3\exp(\frac{x}{128}\ln\frac{283}{3}).
\end{equation}
When training the regression model, we optimize a weighted mean squared error, $E_r$, of the predicted pixel, given by:
\begin{equation}
E_r(\mathbf{y, x}) = -\frac{1}{m} \sum_{i=1}^{m} \bigg[(y_i-x_i)^2(y_i - 64)^2\bigg],
\end{equation}
where $y$ is the ground truth position of $\nu_{\mathrm{max}}$ in pixel coordinates. The weights on this error function penalizes incorrect predictions more when the true $\nu_{\mathrm{max}}$ position is further away from the midpoint of the image (at $z=64$ or $\nu_{\mathrm{max}}\simeq29\mu$Hz), hence it forces the regression model to predict more accurately for stars showing very low or very high $\nu_{\mathrm{max}}$. \edit1{We determine the network structure and parameters for the regression model using a method similar to that of the classifier model, except that here we quantify the regression performance across a dataset using its mean absolute error (MAE), given by:} 
\edit1{\begin{equation}
\mathrm{MAE}(\mathbf{y, x}) = \frac{1}{m} \sum_{i=1}^{m} |y_i-x_i|,
\end{equation}}
\edit1{where $y$ is the pixel-coordinate ground truth and $x$ as the pixel-coordinate predicted value. We use MAE instead of $E_r$ as a metric because by construction, $E_r$ is highly sensitive to outliers and does not weight each prediction equally. Hence, it is not as robust as the MAE in representing the regression performance. The outcome of network structure and parameter selection for the regression model is given in Table \ref{regressiontable}. Specifically, we find that a trained regression model with network structure and parameters as defined in Table \ref{regressiontable} performs the best on the validation set with $\mathrm{MAE_{\mathrm{val}}} \simeq 5.390$. Although we achieve $\mathrm{MAE} \simeq 1.771$ on the training set during training, we find from visual inspection that the larger $\mathrm{MAE_{\mathrm{val}}}$ is due to outliers between the predictions of the regression model and the ground truth for the validation set, where in certain cases the ground truth is in fact incorrect. We also perform an internal validation step to confirm that this difference between training and validation MAE is not due to model overfitting. To do this, we train a model with structures and parameters given in Table \ref{regressiontable} on 15814 out of 19764 stars in our training set and evaluate its performance on the remaining 3950 stars. We find that the MAE evaluated on the 3950 left out stars is comparable to the training MAE, thus ruling out overfitting as the main source of the high $\mathrm{MAE_{\mathrm{val}}}$.
We will explore the regression outliers in greater detail in Section \ref{regperformance}.}

\begin{figure}
	\centering
	\includegraphics[width=\linewidth]{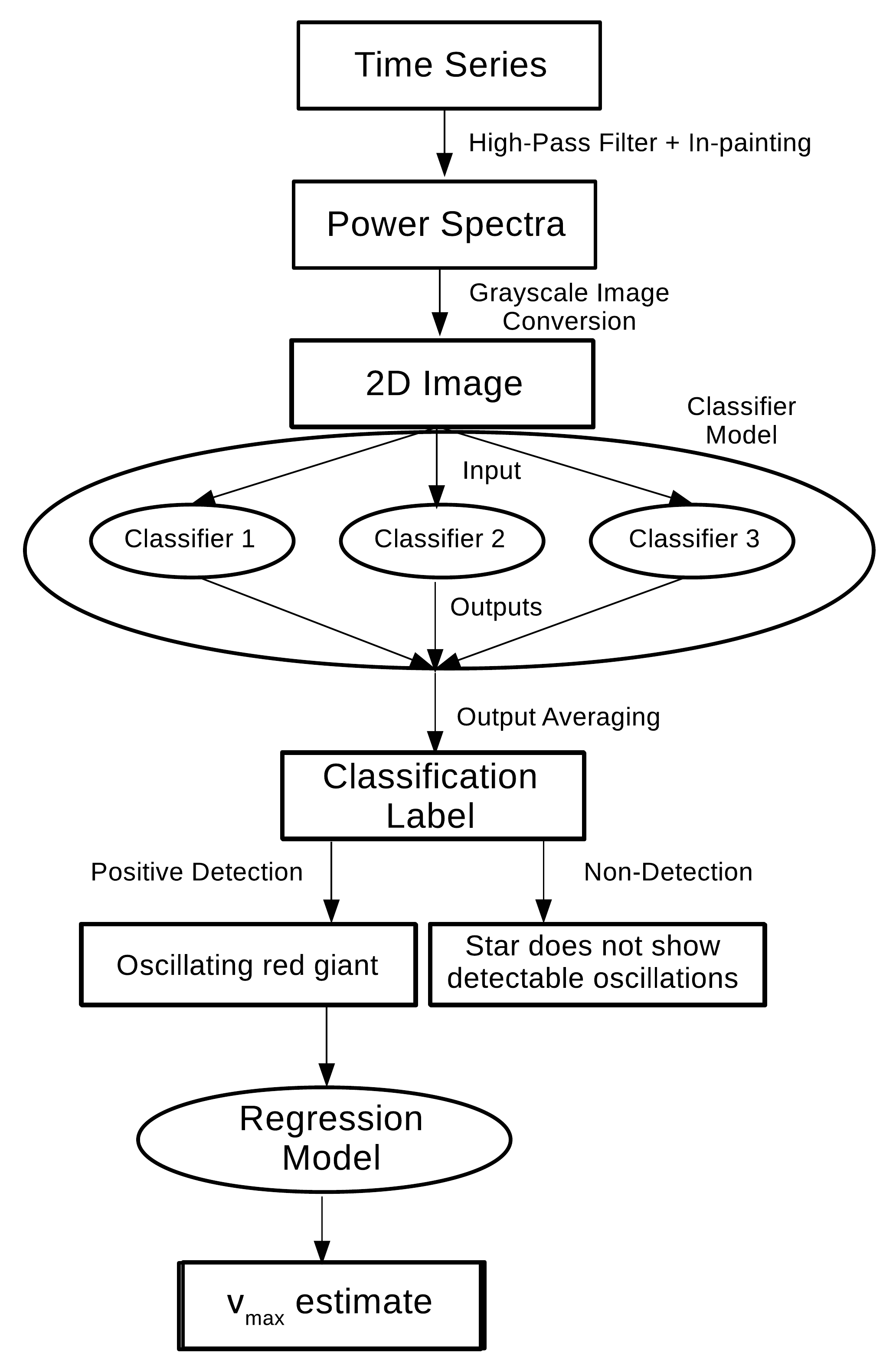}
	\caption{A flowchart of processes involved for the detection of oscillations in this study. Deep learning models are categorized with elliptical shapes, while data objects and products are in rectangles.}
	\label{flowchart}
\end{figure}

We also use Monte Carlo dropout with our $\nu_{\mathrm{max}}$ estimate as the average of 10 forward passes. However, the variance of our predictions is now added with the \textit{inverse model precision}, $\tau^{-1}$ \citep{Gal_2016}, where
\begin{equation}
\tau = \frac{(1-p_\mathrm{drop})l^2}{2m\lambda},
\end{equation}
with $p_\mathrm{drop} =0.375$, $\lambda = 7.5\times10^{-4}$, and $l=5$ as the prior length scale of the data. We use the square root of this modified variance as our prediction uncertainty. A summary of the steps involved in this study is illustrated in the flowchart in Figure \ref{flowchart}. Our deep learning models are trained with the Adam optimizer \citep{Kingma_2014} and constructed with the Keras library \citep{Keras} built on top of Tensorflow \citep{Tensorflow}. Model training utilizes a Quadro K620 GPU and the NVIDIA cuDNN library \citep{Chetlur_2014}. 

\section{Results}

\subsection{Classifier Model Performance}
 We quantify the predictions of the classifier in the form of a \textit{confusion matrix}, which bins the stars according to their predicted class and their truth labels. We tabulate the classifier performance on the validation set (K2 Campaign 6 stars) in Table \ref{c6confusion}. If we denote $c_{i,j}$ as the matrix element on row $i$ and column $j$, the precision of the classifier is defined as $c_{2,2}/(c_{2,1}+c_{2,2})$ (the ratio of the correct positive predictions to all positive predictions) while the recall is defined as $c_{2,2}/(c_{1,2}+c_{2,2})$ (the ratio of the correct positive predictions to all true positives). The classifier accuracy is then $(c_{1,1}+c_{2,2})/(c_{1,1}+c_{2,2}+c_{1,2}+c_{2,1})$ (the ratio of all correct predictions to all predictions). 

\begin{table} 
	\centering
	\caption{Confusion matrix for the validation set comprising stars from K2 Campaign 6. The binary label 0 corresponds to non-detections, while 1 corresponds to a positive detection of solar-like oscillations.  
	}
	\label{c6confusion}

      \begin{tabular}{|c|c|c|c|}
      \cline{3-4}
      \multicolumn{1}{c}{}&\multicolumn{1}{c|}{}&\multicolumn{2}{c|}{Truth Label} \\
      \cline{3-4}
      \multicolumn{1}{c}{}&\multicolumn{1}{c|}{} & 0 & 1\\
      \hline
      \multirow{2}{*}{Predicted Class}& 0 &4401 &52\\
      \cline{2-4}
      & 1 & 84&2659\\

      \hline		
      \end{tabular}

\end{table}

While the validation dataset does not provide an unbiased measure of the classifier's ability to generalize as compared to the test set, it can still be useful in indicating its general performance in detecting oscillating red giants observed by K2. From Table \ref{c6confusion}, the classifier obtains an accuracy of 0.981, a precision of 0.969, and a recall of 0.981. These results are encouraging because it shows that we can perform almost as well as the expert eye when predicting on thousands of stars, though the classifier does it in a matter of seconds. Nonetheless, we investigate the prediction errors of the classifier to better understand its performance. We can see from the table and the precision value that slightly more errors on the validation set come from true non-detections that are predicted as positive detections (\textit{false positives}) compared to true positive detections that are predicted as non-detections (\textit{false negatives}). 

\begin{figure*}
	\centering
	\includegraphics[width=\linewidth]{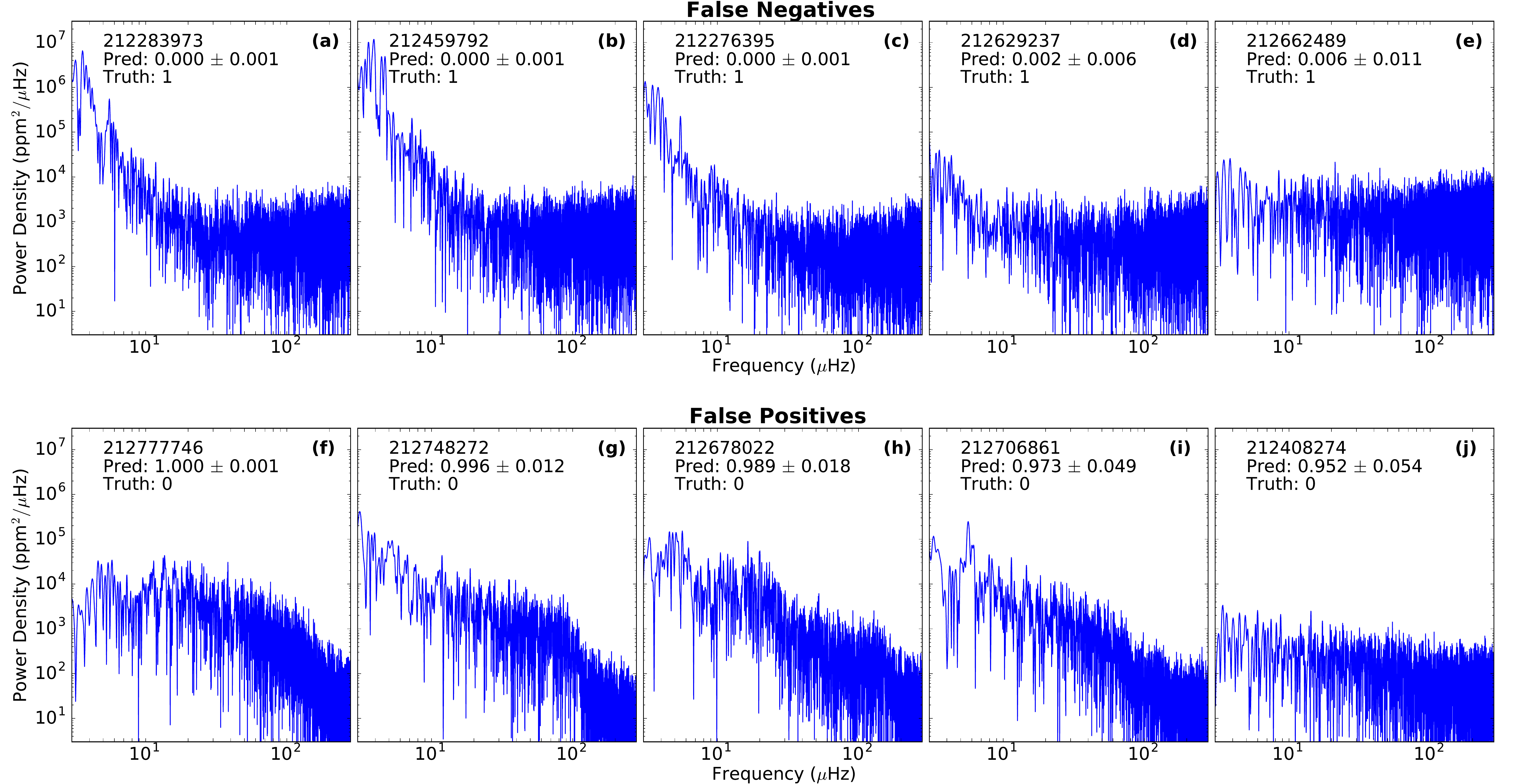}
	\caption{Typical highly disputed false negatives (top row) and false positives (bottom row) of the validation set comprising Campaign 6 stars. The EPIC IDs, prediction probabilities, and the binary ground truth labels for each star are indicated.}
	\label{c6worst}
\end{figure*}
\begin{figure*}
	\centering
	\includegraphics[width=\linewidth]{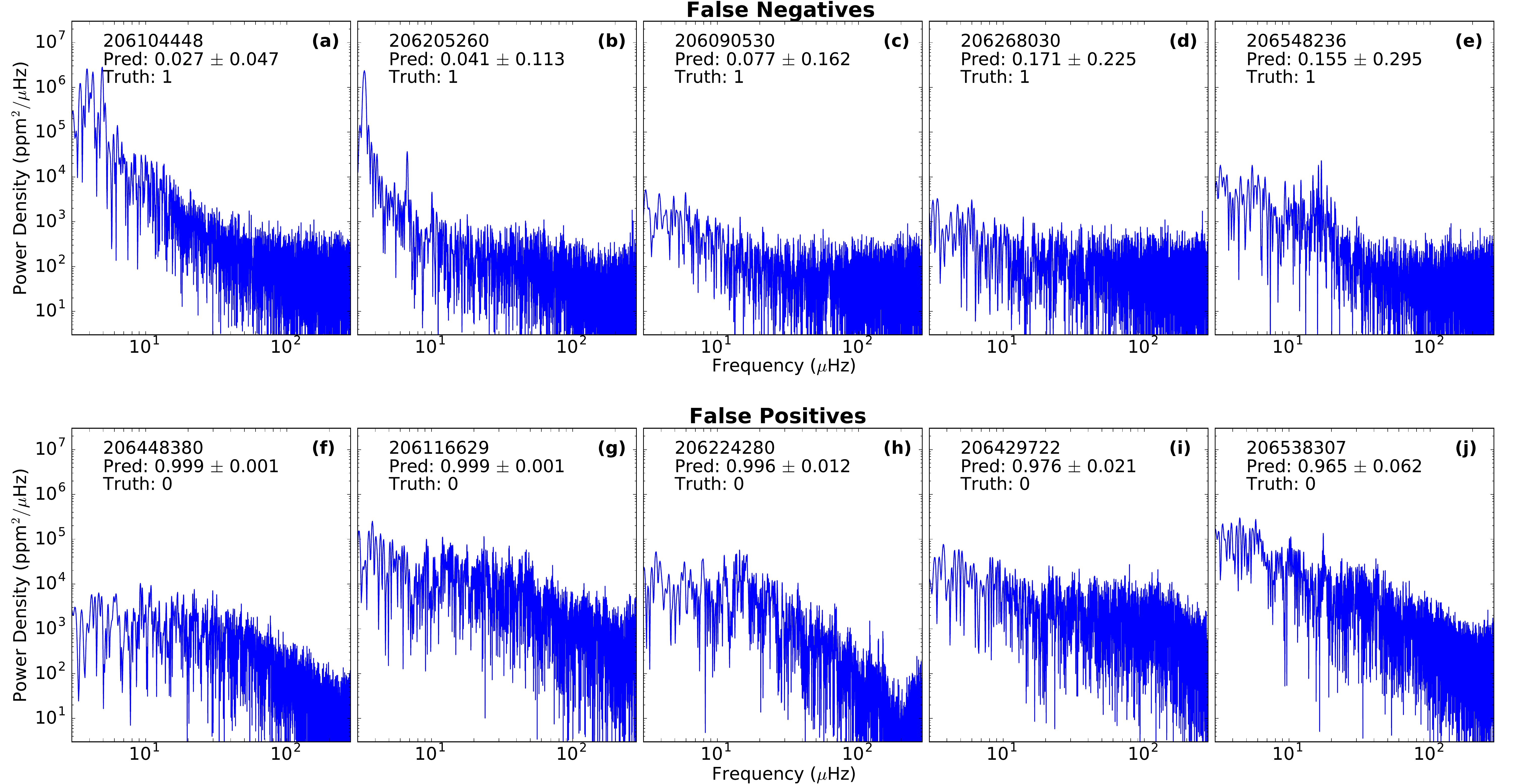}
	\caption{Typical highly disputed false negatives (top row) and false positives (bottom row) of the test set comprising Campaign 3 stars. The EPIC IDs, prediction probabilities, and the binary ground truth labels for each star are indicated.}
	\label{c3worst}
\end{figure*}

For the majority of disputed predictions it is genuinely difficult to determine if the truth label is correct or not, particularly for those with $p$ close to 0.5. There are, however, cases where the classifier is highly confident ($p$ close to 0 or 1) but predicts the opposite classification to the truth label. We identify such cases as highly disputed predictions, in which we can usually verify the classifier's correctness by eye. We present a few typical highly disputed false negatives and false positives from the validation set in Figure \ref{c6worst} so that we can point out possible causes for the disputes. Among the false negatives, we see that the power spectra in Figures \ref{c6worst}a-c potentially have power excesses at $\nu_{\mathrm{max}} \leq5\mu$Hz. However, the low resolution of the power spectra at these frequencies makes it difficult to fully distinguish a `Gaussian-like' profile from the steep slope of the power spectra at low frequencies. Furthermore, the `elbow'-like profile of the spectra makes them appear similar to the non-detections with $\nu_{\mathrm{max}}$ below the plotted (and resolved) range (see Figure \ref{comparisons}h). Meanwhile, Figures \ref{c6worst}d-e show spectra with a relatively flat profile. Their truth label as positive detection possibly relies on the assumption that they could be heavily blended (hence low granulation and oscillation power) low-$\nu_{\mathrm{max}}$ stars.  We regard the labels of these stars as ambiguous. In fact, we see that the presented false negatives share a white noise-dominated profile at higher frequencies with similar power, which is very likely to be a feature used by the classifier to decide the output label. In contrast, we see that the false positives (Figures \ref{c6worst}f-j) share a downward sloping profile, while sometimes having a `bump' along the power spectra that vaguely resembles a `Gaussian-like' power excess (see Figure \ref{c6worst}h, $\nu \simeq 105\mu$Hz). We note that in most highly disputed false positives, the classifier confuses a rather unusual power spectrum profile for one that shows granulation and oscillations. 


We show the classifier performance on the test set (Campaign 3 stars) in Table \ref{c3confusion}. The classifier obtains an accuracy of 0.991, a precision of 0.988, and a recall of 0.934. The relatively low recall value with respect to precision implies that the classifier predicts many more false negatives compared to false positives. This is the opposite scenario to the validation set. We show a few typical highly disputed false positives and false negatives from the test set in Figure \ref{c3worst}. The false negatives in Figures \ref{c3worst}a-b show low $\nu_{\mathrm{max}}$ power spectra profiles similar to that of Figures \ref{c6worst}a-c, while those in Figures \ref{c3worst}c-d show relatively flat profiles. As for Figures \ref{c6worst}d-e, the correct truth labels for these flat profile stars are ambiguous. It is possible that oscillations can be seen from a smoothed version of the power spectra, which is currently not provided to the classifier, but we will investigate its use in future work. 
Figure \ref{c3worst}e clearly shows a feature resembling  oscillation excess at $\nu \simeq15\mu$Hz, however the profile becomes flat at higher frequencies because of the high level of white noise. It is likely that the lack of a typical granulation slope causes the classifier to determine this star as a non-detection, although with a high prediction uncertainty. The highly disputed false positives in the test set (Figures \ref{c3worst}f-j) generally show profiles with distinct granulation slopes, but without a clear oscillation excess feature, similar to the highly disputed false positives on the validation set. Though we can understand what types of mistakes are performed by the classifier from observing typical highly disputed predictions in both validation and test sets, we can better understand why this happens by probing what the classifier `sees', which we investigate in Section \ref{visualization}.

\edit1{We summarize the classifier performance statistics from this section in Table \ref{performancesummary}, where we notably observe that the test set accuracy is higher than the validation set accuracy. In addition, we see that the source of most misclassified stars across different campaigns are not consistently due to false positives or false negatives. We attribute these results to the differences in the target samples and possibly in the instrumental features of the data between different campaigns. Nonetheless, the classifier still shows a high classification accuracy across both campaigns.}

\begin{table} 
	\centering
	\caption{Confusion matrix for the \edit1{test} set comprising stars from Campaign 3 of K2.  
	}
	\label{c3confusion}

    \begin{tabular}{|c|c|c|c|}
    \cline{3-4}
    \multicolumn{1}{c}{}&\multicolumn{1}{c|}{}&\multicolumn{2}{c|}{Truth Label} \\
    \cline{3-4}
    \multicolumn{1}{c}{}&\multicolumn{1}{c|}{} & 0 & 1\\
    \hline
    \multirow{2}{*}{Predicted Class}& 0 &11397 &98\\
    \cline{2-4}
    & 1 & 16&1354\\

    \hline

    \end{tabular}

\end{table}

\begin{table} 
	\centering
	\caption{Summary statistics for the classifier on the validation and test sets.  
	}
	\label{performancesummary}

    \begin{tabular}{|c|c|c|}
    \cline{2-3}
    \multicolumn{1}{c|}{}&Campaign 6 (Validation)&Campaign 3 (Test)\\
    \hline
    Accuracy&98.1\% & 99.1\% \\
    \hline
    Precision&96.9\% & 98.8\%\\
    \hline
    Recall & 98.1\% & 93.3\%\\

    \hline

    \end{tabular}

\end{table}

\begin{figure*}[h!]
	\centering
	\includegraphics[width=\linewidth]{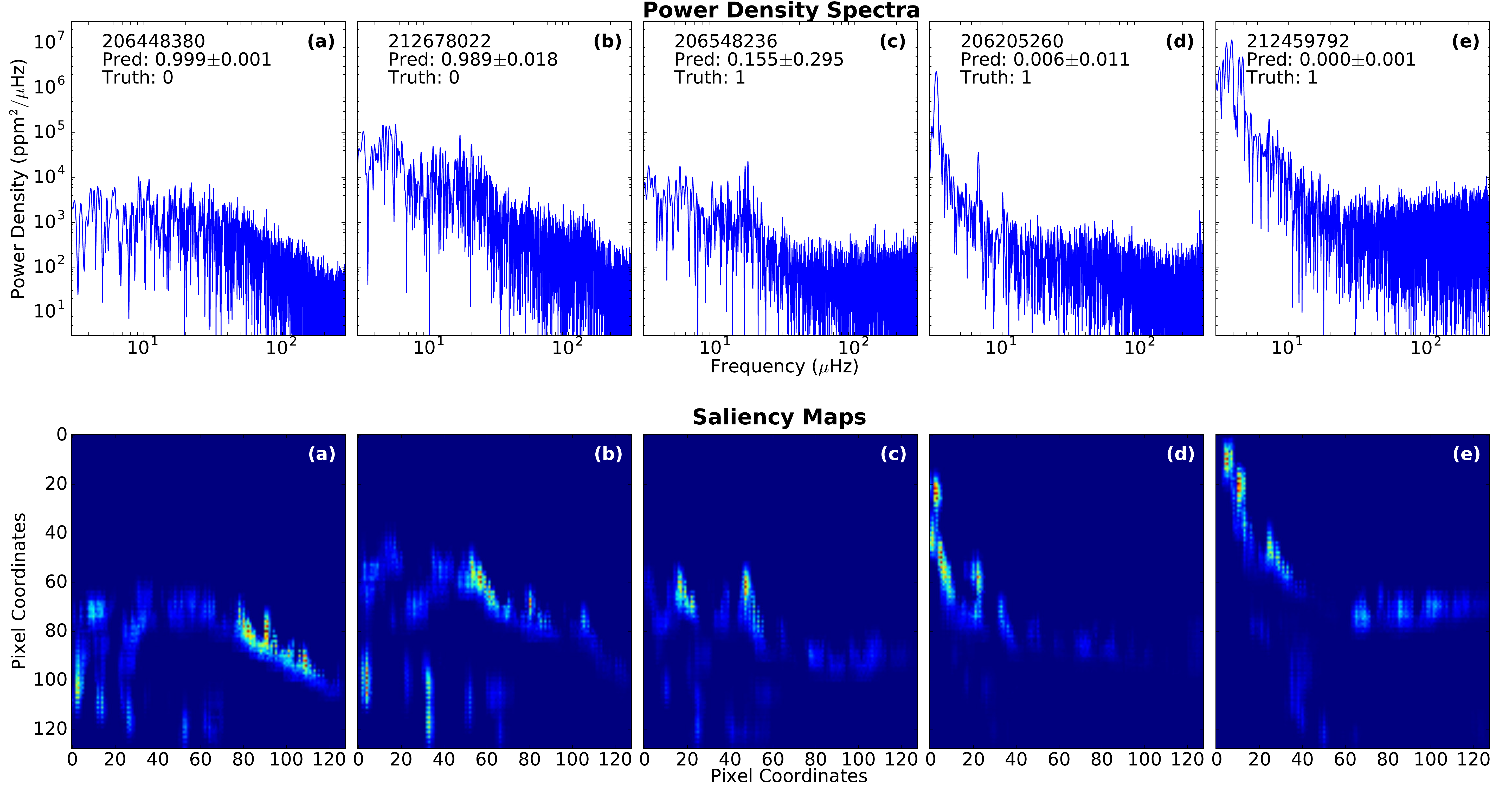}
	\caption{Saliency maps for a subset of highly disputed false positives and false negatives from Figures \ref{c6worst} and \ref{c3worst}. (a-b) are predicted as positive detections, hence their corresponding heatmaps show regions that contribute towards a \textit{positive detection}, while heatmaps corresponding to (c-e) show regions that contribute towards a \textit{non-detection} because they are predicted as non-detections.}
	\label{saliencyworst}
\end{figure*}

\subsection{Classifier Model Visualizations}\label{visualization}

While convolutional neural networks are commonly described as black boxes, there exist methods that allow us to carry out `reverse engineering' by visualizing the extracted features in the network. This enables us to better understand the performance of our trained deep learning models. An advantage of working in the 2D domain is that these visualizations form images that are easy to comprehend by the viewer. In this paper, we use two different methods of visualizing our deep learning models.

The first method is by constructing a \textit{saliency map} \citep{Simonyan_2013}. A saliency map allows us to identify features within the image that contribute the most to the output prediction probability of the deep learning network. In other words, it visualizes the regions where the network gives the most `attention' in assigning a class label to a particular image. For example, if a star is predicted as a positive detection by the classifer, the saliency map shows the regions that contributed to the positive detection, while for stars predicted as non-detections, the map shows the regions that contributed to the non-detection classification. The saliency map is constructed by computing the gradient of the predicted output class with respect to each pixel in the input image by backpropagation. The magnitude of each pixel's gradient shows the sensitivity of the network output with respect to small changes in values of that pixel, hence it highlights image features that are most important (salient) for the deep learning model.
\begin{figure*}
	\centering
	\includegraphics[width=\linewidth]{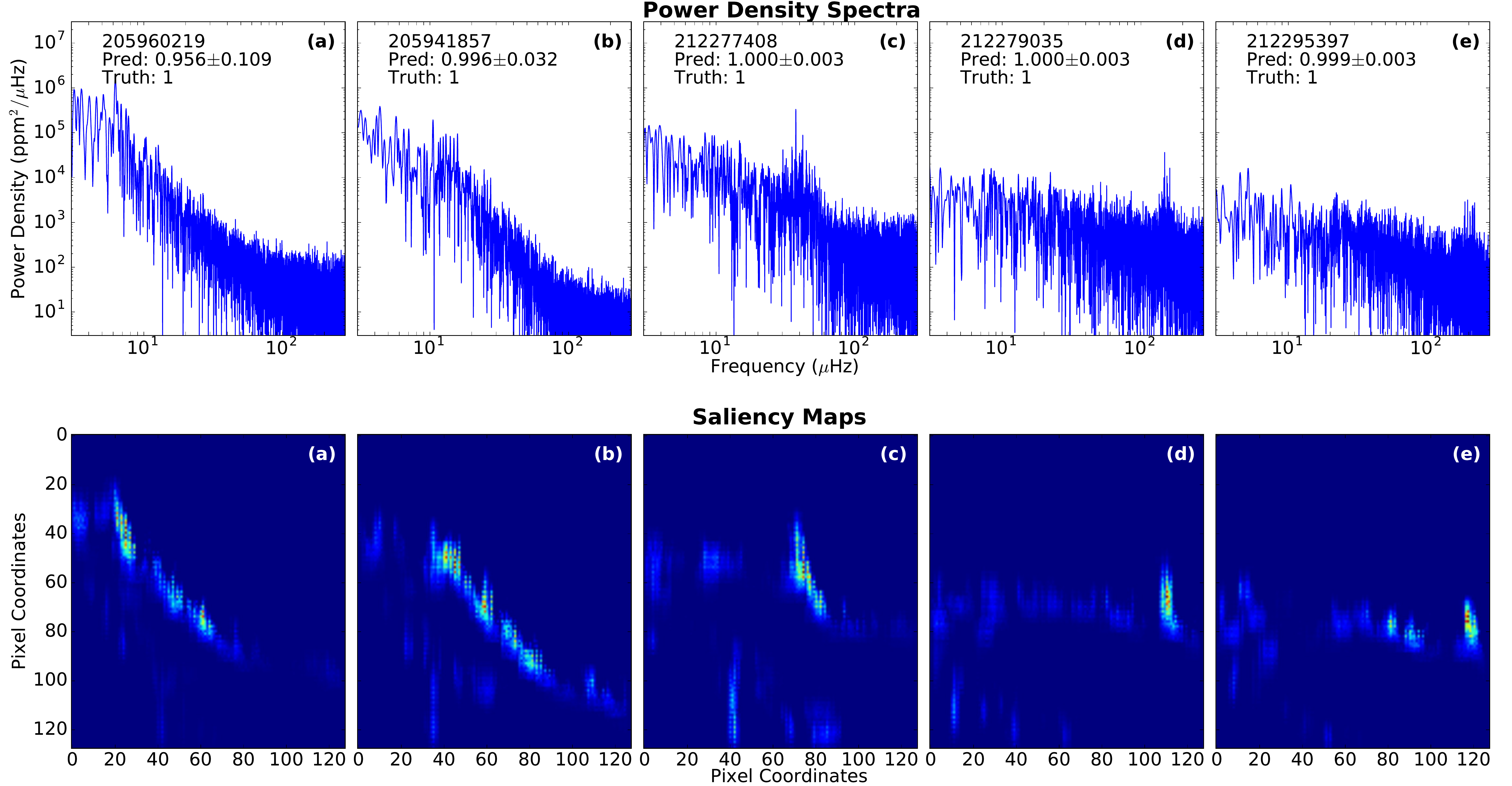}
	\caption{Power density spectra of oscillating red giants observed by K2 (top row), with their corresponding saliency maps (bottom row). All the listed stars are classified as positive detections. Regions with hotter colours in the heatmaps are given greater `attention' by the classifier, and contribute highly towards a \textit{positive detection}. Stars in panels a-b are from Campaign 3, and  the rest are from Campaign 6.}
	\label{saliencydetection}
\end{figure*}
\begin{figure*}
	\centering
	\includegraphics[width=\linewidth]{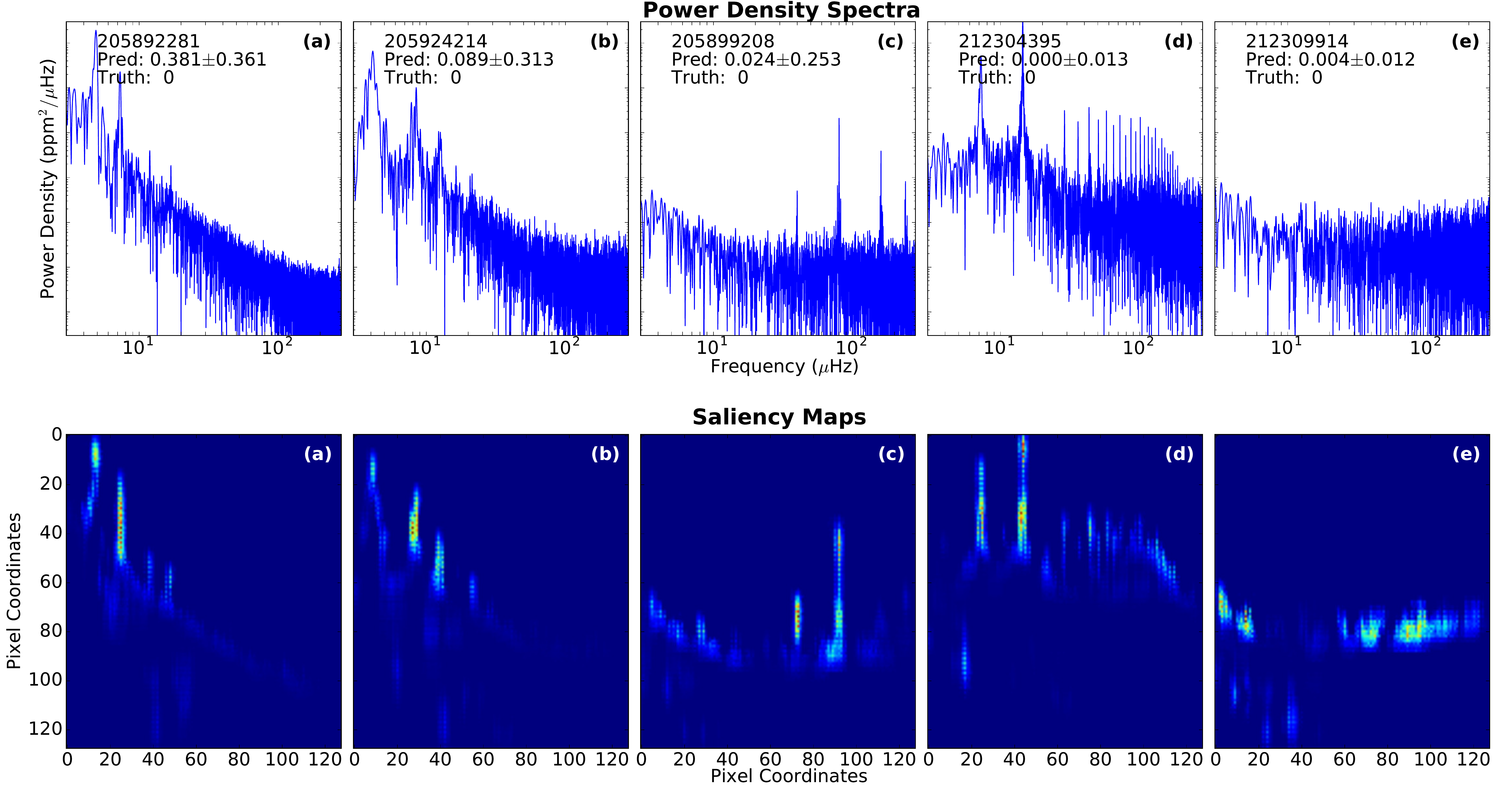}
	\caption{Power density spectra of stars observed by K2 that do not show solar-like oscillations (top row), with their corresponding saliency maps (bottom row). Regions with hotter colours in the heatmaps are given greater `attention' by the classifier, and highly contribute towards a \textit{non-detection}. Stars in plots a-b are from Campaign 3, with the rest from Campaign 6.}
	\label{saliencynondetection}
\end{figure*}

The second method of visualization is by \textit{activity maximization} \citep{Erhan_2009, Simonyan_2013}. This method numerically generates an image that maximizes the activations for each filter within a layer. In other words, it shows the features which will generate the largest output for each filter within a layer. Hence, visualizing the activity maximization at each convolutional layer in the network reveals the form of image patterns that each template-matching filter learns to look for when predicting on an image, while if performed at the output layer, it shows a representative image of a specific class. These activity-maximizing images are computed by optimizing a loss function that penalizes small filter activations with respect to the input image using gradient descent. The saliency map and activity maximization visualizations in our study are constructed using the Keras Visualization Toolkit \citep{kerasvis}.

\subsubsection{Saliency Maps}\label{saliency}

To understand the decision-making behind a few typical highly disputed predictions in Figures \ref{c6worst} and \ref{c3worst}, we construct their saliency maps in Figure \ref{saliencyworst}. By observing regions of the image that highly influence the classifier output (`hotter' colours or stronger highlights on heatmap), we can see that the classifier mostly observes features on the upper portions of the power density profile, similar to the expert eye. The saliency maps show features that contribute towards a prediction for a particular class, hence the maps in Figures \ref{saliencyworst}a-b (false positives) show regions in the image that contribute towards a positive detection. For the false positive in Figure \ref{saliencyworst}a , we see that the classifier pays particular attention to the sloping region of the power spectrum despite no presence of a clear power excess while in Figure \ref{saliencyworst}b, the main highlighted regions reveals features that the classifier mistakes to be the oscillation excess.

The maps in Figures \ref{saliencyworst}c-e (false negatives) show regions that contribute towards a non-detection. Interestingly, they mainly highlight \edit1{the presence of multiple} sharp vertical peaks at low frequencies. While these could be representative of non-detections showing low frequency variability of other types, using them as a classification criterion is not \edit1{reliable} because the power excess of oscillating red giants can potentially also form sharp vertical peaks at low frequencies in the power spectrum profile as well (for example $\nu_{\mathrm{max}}\simeq15\mu$Hz for Figure \ref{saliencyworst}c and $\nu_{\mathrm{max}}\simeq3\mu$Hz for Figure \ref{saliencyworst}e). Thus, the inaccuracy in Figure \ref{saliencyworst}e is not due the failure of the classifier to detect the supposed oscillation excess, but rather that it recognized it as a non-detection feature \edit1{in the presence of the other sharp peak at a lower frequency}. We also note that there are faint highlights for regions along the flat white noise profile in Figure \ref{saliencyworst}e, implying that the classifier attributes certain features from the white noise to a non-detection as well.

To further examine how the `attention' of the classifier is distributed, we also show examples of correctly identified stars. First, we present the saliency maps of oscillating giants in Figure \ref{saliencydetection}. In each of these maps, strong highlights of the power excess indicate that the classifier successfully identifies the envelope containing oscillation modes as a feature that highly impacts its prediction output. Furthermore, in the saliency maps of Figures \ref{saliencydetection}a and \ref{saliencydetection}b, highlighted regions along the descending slope of the power density profile leads us to infer that the presence of a steep granulation slope is an important feature for a positive detection.

We show examples of correctly identified non-detections with their saliency maps in Figure \ref{saliencynondetection}. These maps show features that contribute highly towards a non-detection, hence we can see in Figures \ref{saliencynondetection}a-d that sharp vertical peaks within the power spectrum are considered as non-detection features. 
Finally, Figure \ref{saliencynondetection}e shows a spectrum dominated by white noise, and we see from the saliency map that the classifier attributes flat noise features at very low ($\nu\apprle5\mu$Hz) and mid-range frequencies ($30\mu$Hz $\apprle\nu\apprle100 \mu$Hz) to a non-detection. This can also be seen in Figure \ref{saliencyworst}e.

\subsubsection{Activity Maximization}
\begin{figure}
	\centering
	\includegraphics[width=\linewidth]{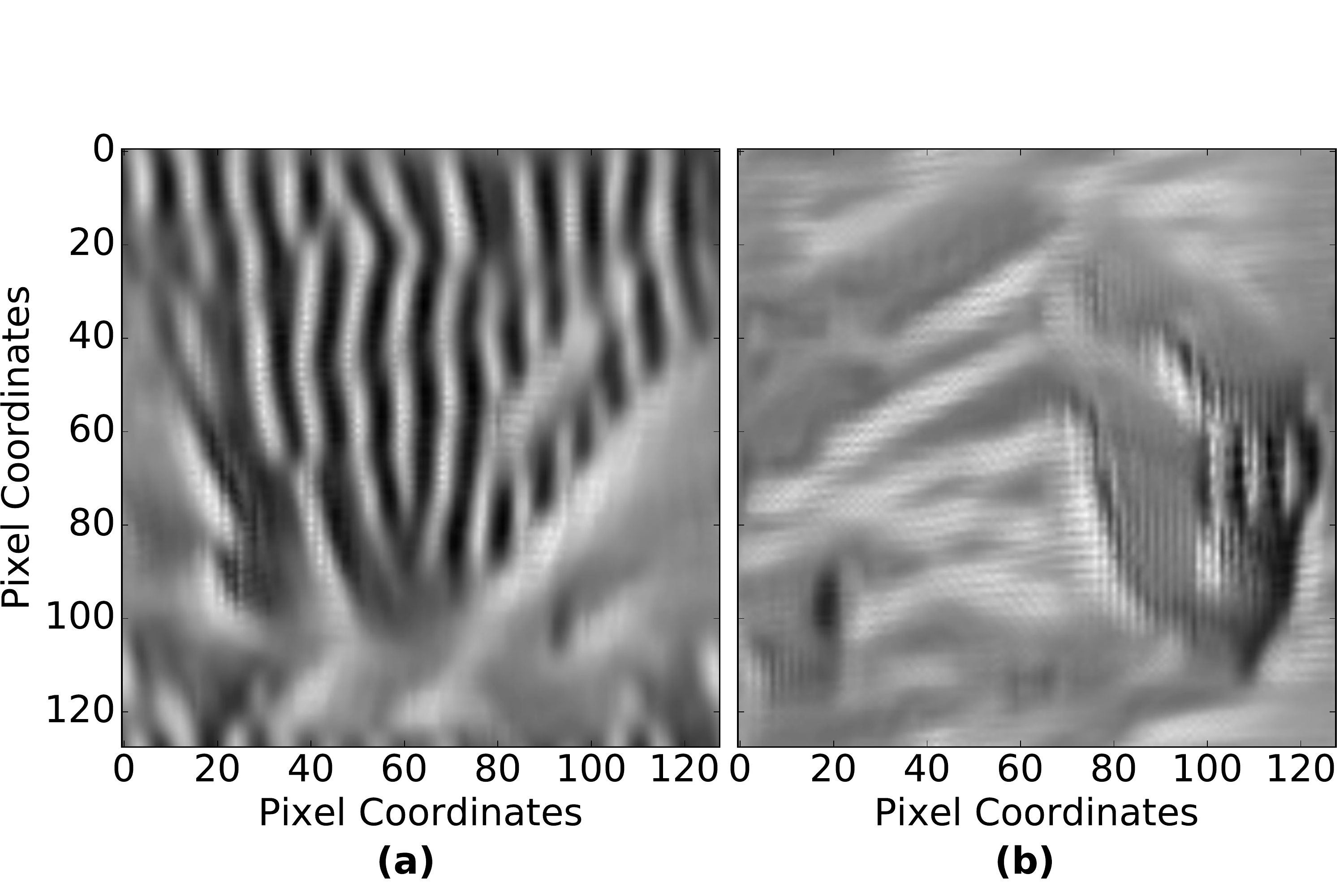}
	\caption{Generated images that are representative of a (a) non-detection and a (b) positive detection by the classifier model. White grooves in the image represent features that are shown by the power spectrum profile.}
	\label{activitymax}
\end{figure}
The type of features that contribute towards predicting a particular class can also be visualized by generating class-representative images from the classifier, which we show in Figure \ref{activitymax}. The representative image of a non-detection (Figure \ref{activitymax}a) shows a very distinct vertical groove pattern throughout the upper half of the image, potentially corresponding to \edit1{multiple} sharply defined vertical peaks in those regions. If this is the case, it would strongly agree with our observations in Figures \ref{saliencyworst} and \ref{saliencynondetection} regarding \edit1{multiple} sharp, vertical peaks at low to mid-range frequencies being attributed to non-detections. As a result of the classifier's rule for non-detection, it is possible that there exists a detection bias against highly luminous stars oscillating at very low frequencies ($\nu_{\mathrm{max}}\apprle5\mu$Hz). The lack of frequency resolution at such a low frequency range can produce sharply defined features in the log-log plot of the power spectrum (e.g. Figure \ref{saliencyworst}e), which is then potentially attributed by the classifier to a non-detection. We note, however, that the vertical groove pattern does not extend into the lower right section of the image. Instead, the vertical pattern at this location can be seen in the representative image of a positive detection (Figure \ref{activitymax}b), which is where we would expect the peaks from oscillation excess with $\nu_{\mathrm{max}}\apprge100\mu$Hz to be located. Oscillation power excess at such frequencies will typically be sitting on a flatter power density profile and have the appearance of a sharp peak (see Figures \ref{comparisons}c-d). Hence, we infer that a single strong power excess hump at those frequencies are likely to be attributed to a positive detection instead of a non-detection. Another noticeable pattern within Figure \ref{activitymax}b is the presence of white angular grooves at the image center. These grooves strongly resemble the profile of a power spectrum having a granulation slope on which sits a distinct power excess (see Figures \ref{granulation} and \ref{comparisons}b). The presence of the grooves at different heights in the image also indicate that the classifier learns to look for the pattern at different power density levels.

\begin{figure*}
	\centering
	\includegraphics[width=\linewidth]{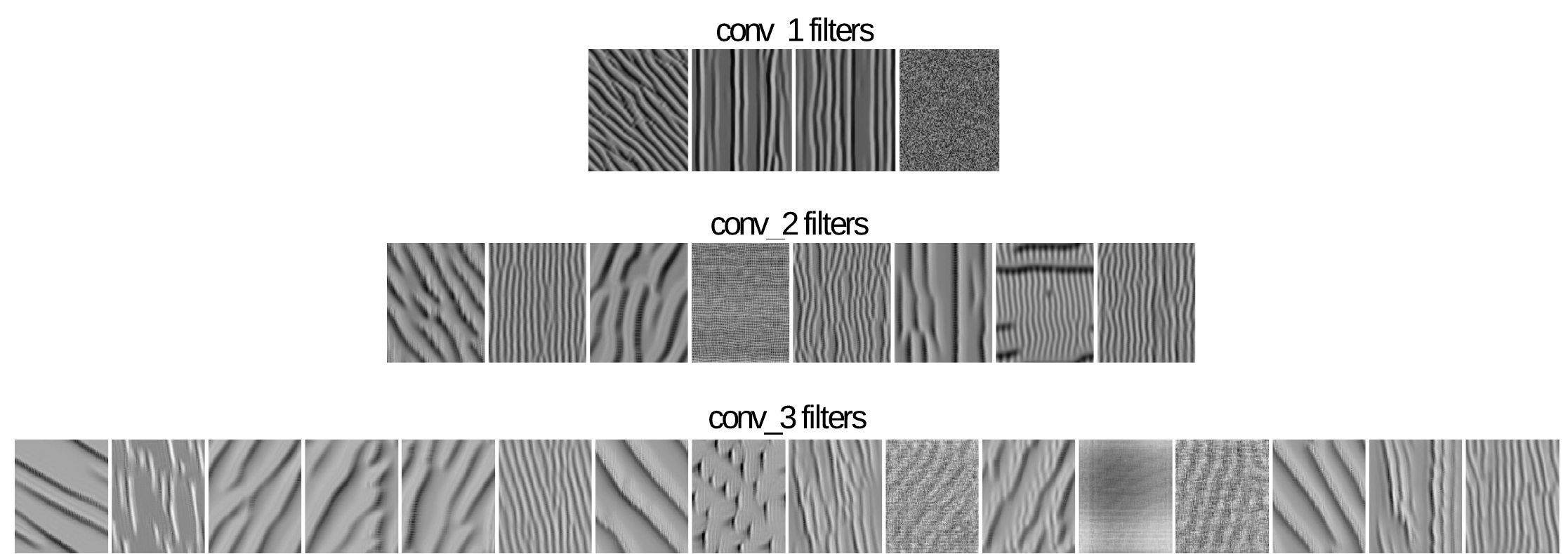}
	\caption{Visualization of the convolutional filters at each layer within the classifier model by activity maximization. The patterns in each filter define the learned template the network uses for convolution. Note that the patterns further down the network become increasingly abstract, corresponding to learned higher level features.}
	\label{classificationconvfilters}
\end{figure*}

While the images in Figure \ref{activitymax} may be generally difficult to interpret, the activity-maximized convolutional filters in Figure \ref{classificationconvfilters} clearly show the image patterns of interest for classification. From the filters showing repeating diagonal grooves, we see that the classifier learns to detect the presence of upward and downward slopes at different parts of the image. The composite of these slopes potentially form the angular groove pattern, which as discussed previously, is attributed to the shape of an oscillation power excess on top of granulation. Interestingly, the 8th filter in the conv\_3 layer (8th image from left in the bottom row in Figure \ref{classificationconvfilters}) even shows sharp triangular patterns resembling the shape of a smoothed power excess. There are also filters showing vertical stripes, which we infer are dedicated to detecting the presence of non-detection features as illustrated in Figure \ref{activitymax}a. 

\subsection{Regression Model Performance}\label{regperformance}
\begin{figure}
	\centering
	\includegraphics[width=\linewidth]{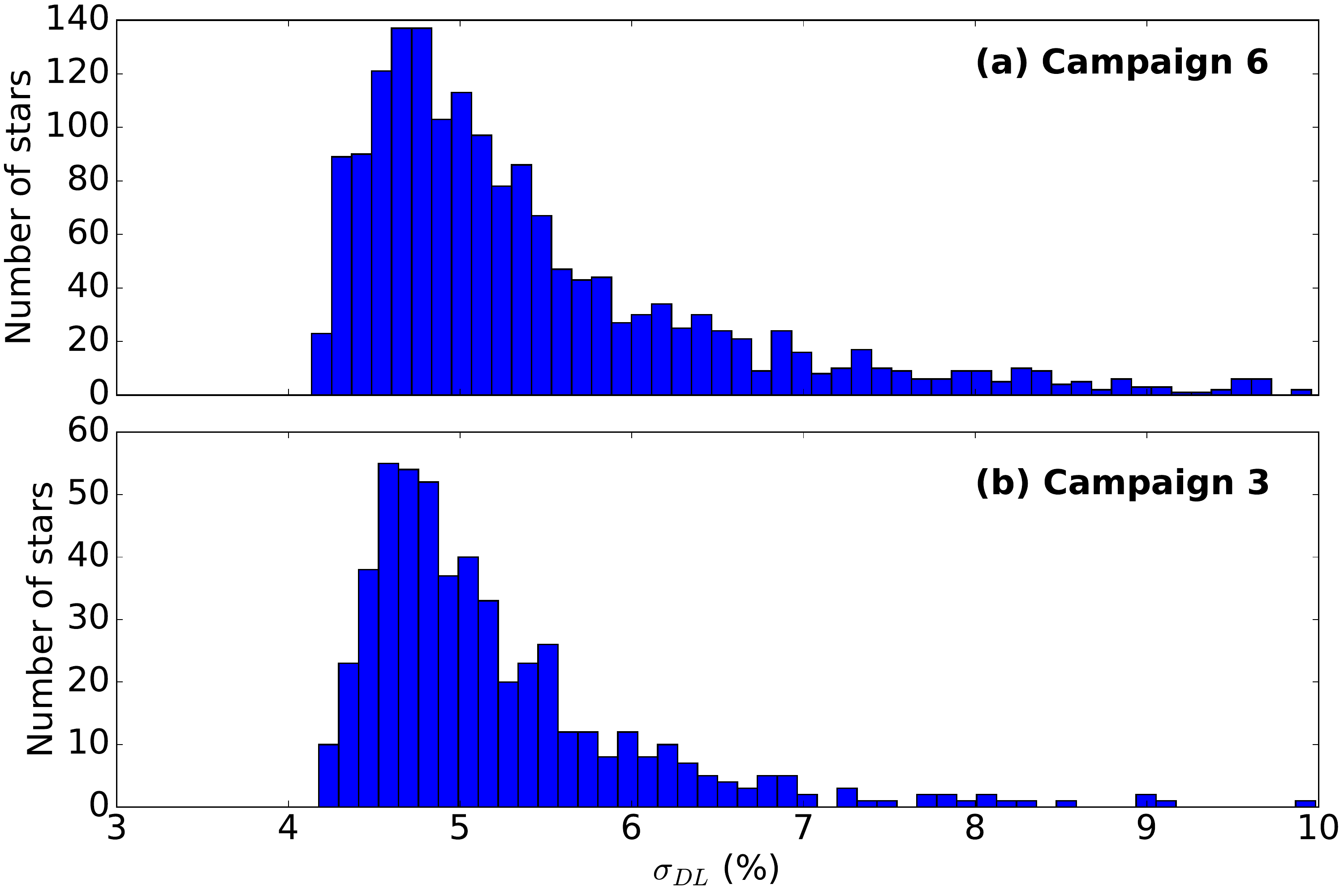}
	\caption{The distribution of fractional uncertainties of the regression model's predictions on (a) Campaign 6 and (b) Campaign 3 stars.}
	\label{sigmadl}
\end{figure}
\begin{figure}
	\centering
	\includegraphics[width=\linewidth]{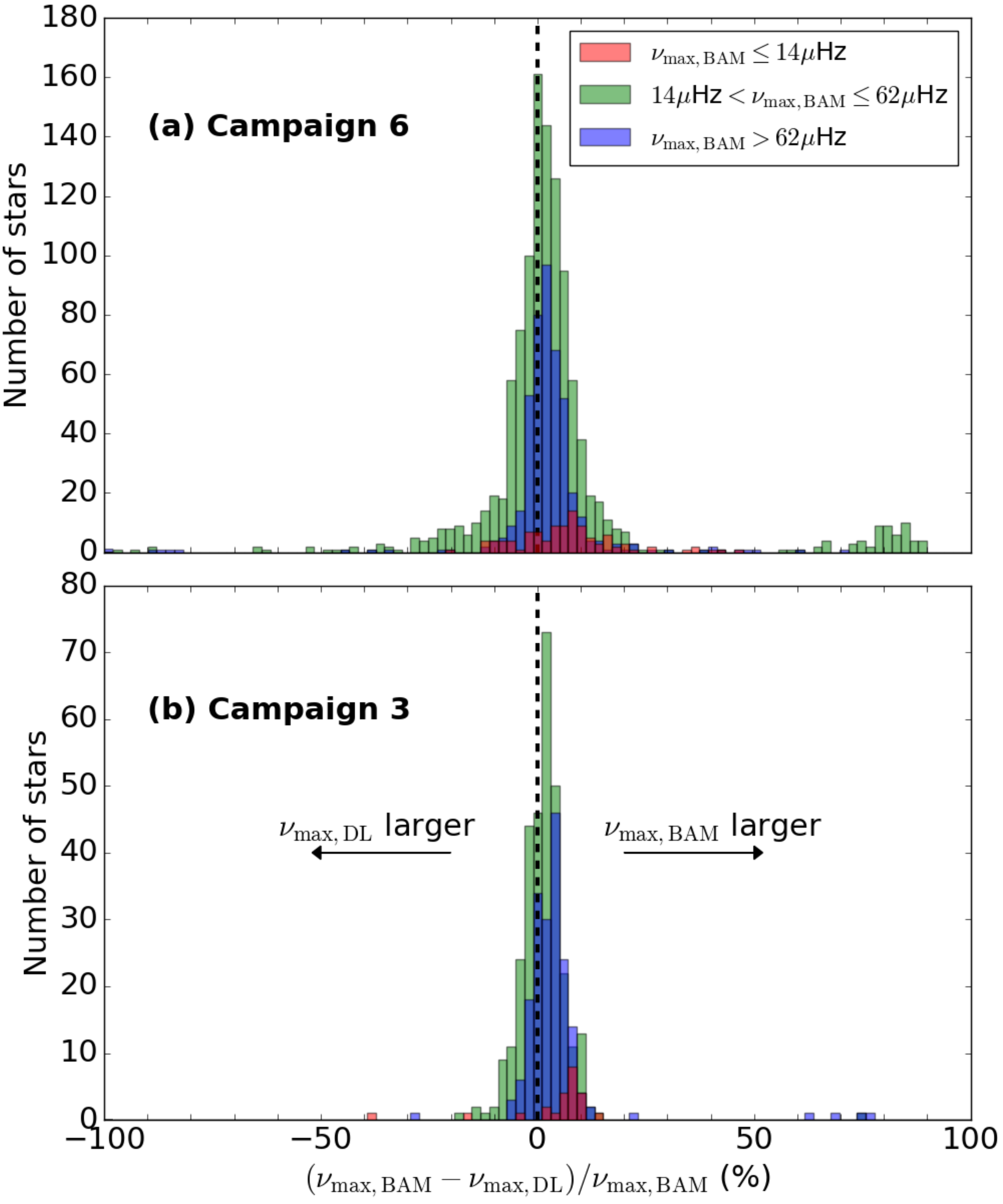}
	\caption{Fractional differences between measured $\nu_{\mathrm{max}}$ values from the BAM pipeline, $\nu_{\mathrm{max, BAM}}$, and from the deep learning model, $\nu_{\mathrm{max, DL}}$, for (a) Campaign 6 stars (validation set) and (b) Campaign 3 stars (test set). Stars with $\nu_{\mathrm{max, BAM}}$ in the range of pixel x-coordinates 0-43 (first one-third portion of image) are in red, while 43-85 (second one-third portion) are in green and 85-128 (final one-third portion) are in blue. The vertical dashed line marks the origin to guide the eye.}
	\label{regressionhist}
\end{figure}

We show the distribution of fractional uncertainties for the regression model, $\sigma_{DL}$, in Figure \ref{sigmadl}, where we find typical values of about 5\% for both Campaigns 3 and 6. We further evaluate the performance of the regression model by measuring the fractional differences of its predicted $\nu_{\mathrm{max}}$ values, $\nu_{\mathrm{max, DL}}$, with that measured by the BAM pipeline, $\nu_{\mathrm{max, BAM}}$, as shown in Figure \ref{regressionhist} for K2 Campaigns 3 and 6. Near the origin, the fractional differences for each Campaign are approximately normally distributed, with a slight tendency that $\nu_{\mathrm{max, DL}}$ is smaller than $\nu_{\mathrm{max, BAM}}$ on average. We find that 76\% and 94\% of $\nu_{\mathrm{max, DL}}$ predictions are within $\pm10\%$ of $\nu_{\mathrm{max, BAM}}$ for Campaigns 6 and 3, respectively. Moreover, 54\% and 71\% of $\nu_{\mathrm{max, DL}}$ predictions are within $\pm5\%$ of $\nu_{\mathrm{max, BAM}}$ for Campaigns 6 and 3, respectively, hence we infer that the standard deviation of $\nu_{\mathrm{max, DL}}$ from $\nu_{\mathrm{max, BAM}}$ is about 5-7\%. In support of this, we note that BAM's measured $\nu_{\mathrm{max}}$ uncertainties, $\sigma_{\mathrm{BAM}}$, is typically around 2-3\% \citep{Stello_2017}, which in combination with $\sigma_{DL}\simeq5\%$ results in a value consistent with our 5-7\% estimate above.

\begin{figure*}[ht!]
	\centering
	\includegraphics[width=\linewidth]{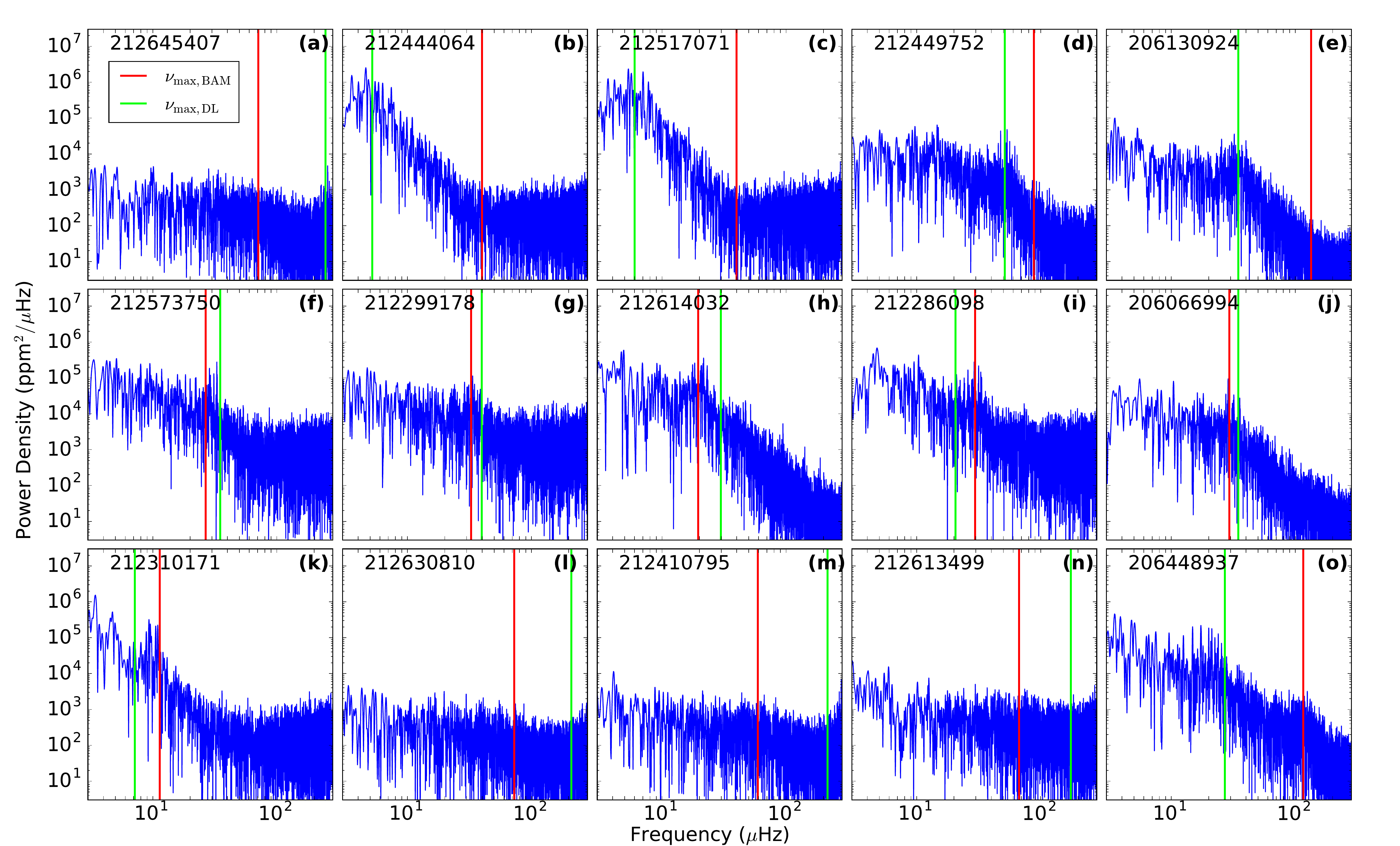}
	\caption{Examples of stars with more than a 20\% fractional difference between $\nu_{\mathrm{max, BAM}}$ and $\nu_{\mathrm{max, DL}}$ in Campaigns 3 and 6. (a-e) $\nu_{\mathrm{max, DL}}$ provides the correct estimate for $\nu_{\mathrm{max}}$. (f-j) $\nu_{\mathrm{max, BAM}}$ provides the correct estimate for $\nu_{\mathrm{max}}$. (k-n) Neither $\nu_{\mathrm{max, BAM}}$ nor $\nu_{\mathrm{max, DL}}$ provides an accurate estimate of the true $\nu_{\mathrm{max}}$. (o) The power density spectra has an unusual profile, such that the true $\nu_{\mathrm{max}}$ cannot easily be determined visually.}
	\label{regoutliers}
\end{figure*}

When measuring how close $\nu_{\mathrm{max, DL}}$ predictions are to $\nu_{\mathrm{max, BAM}}$ with respect to $\sigma_{\mathrm{BAM}}$ on Campaigns 6 and 3, we find that 22\% of our predictions for both Campaigns lie within 1$\sigma_{\mathrm{BAM}}$, while 39\% and 44\% lie within 2$\sigma_{\mathrm{BAM}}$, respectively. These proportions are comparable to the performance of $\nu_{\mathrm{max}}$ estimates assigned by the expert eye in Campaign 1 \citep{Stello_2017}, where 23\% of visual estimates lie within 1$\sigma_{\mathrm{BAM}}$ and 42\% of estimates lie within 2$\sigma_{\mathrm{BAM}}$. This remarkable level of agreement shows that our $\nu_{\mathrm{max, DL}}$ predictions are comparable with human-level visual $\nu_{\mathrm{max}}$ estimates on average, and can be used to estimate fairly accurate $\log g$ values. Our results will also be extremely useful input for subsequent analyses that aims to determine more statistically robust values of $\nu_{\mathrm{max}}$, $\Delta\nu$, and other seismic and granulation properties using parametric model fitting techniques.  

While the vast majority of our $\nu_{\mathrm{max, DL}}$ predictions fall close to $\nu_{\mathrm{max, BAM}}$, Figure \ref{regressionhist} does show some clear outliers. Most outliers seem to group at fractional differences of approximately +75 to +80\%. Additionally, there are a few outliers with a fractional difference above 100\%, however we do not include them in Figure \ref{regressionhist} due to their scarcity. We find that 229 out of 1747 predictions in Campaign 6 have fractional difference magnitudes greater than 20\%, where we have visually verified that $\nu_{\mathrm{max, DL}}$ is the correct $\nu_{\mathrm{max}}$ estimate for 163 cases, while $\nu_{\mathrm{max, BAM}}$ provides the correct estimate for 18 cases, while neither are correct for the remaining 48. For Campaign 3, only 9 out of 525 predictions have fractional difference magnitudes greater than 20\%, where we have determined that $\nu_{\mathrm{max, DL}}$ is correct for 6 cases, while $\nu_{\mathrm{max, BAM}}$ is correct for 1, and neither are correct for the remaining 2.

We show examples of each outlier scenario in Figure \ref{regoutliers}. The top row shows cases where $\nu_{\mathrm{max, DL}}$ (green) is the correct $\nu_{\mathrm{max}}$ estimate, which generally fall into two types. The first type is where the true $\nu_{\mathrm{max}}$ is at higher frequencies, which is successfully detected by the regression model but not by BAM (Figure \ref{regoutliers}a). The second is where BAM predicts $\nu_{\mathrm{max, BAM}}$ (red) to be at the bottom of the granulation curve instead of the location of the power excess (Figures \ref{regoutliers}b-e). 
\begin{figure*}
	\centering
	\includegraphics[width=\linewidth]{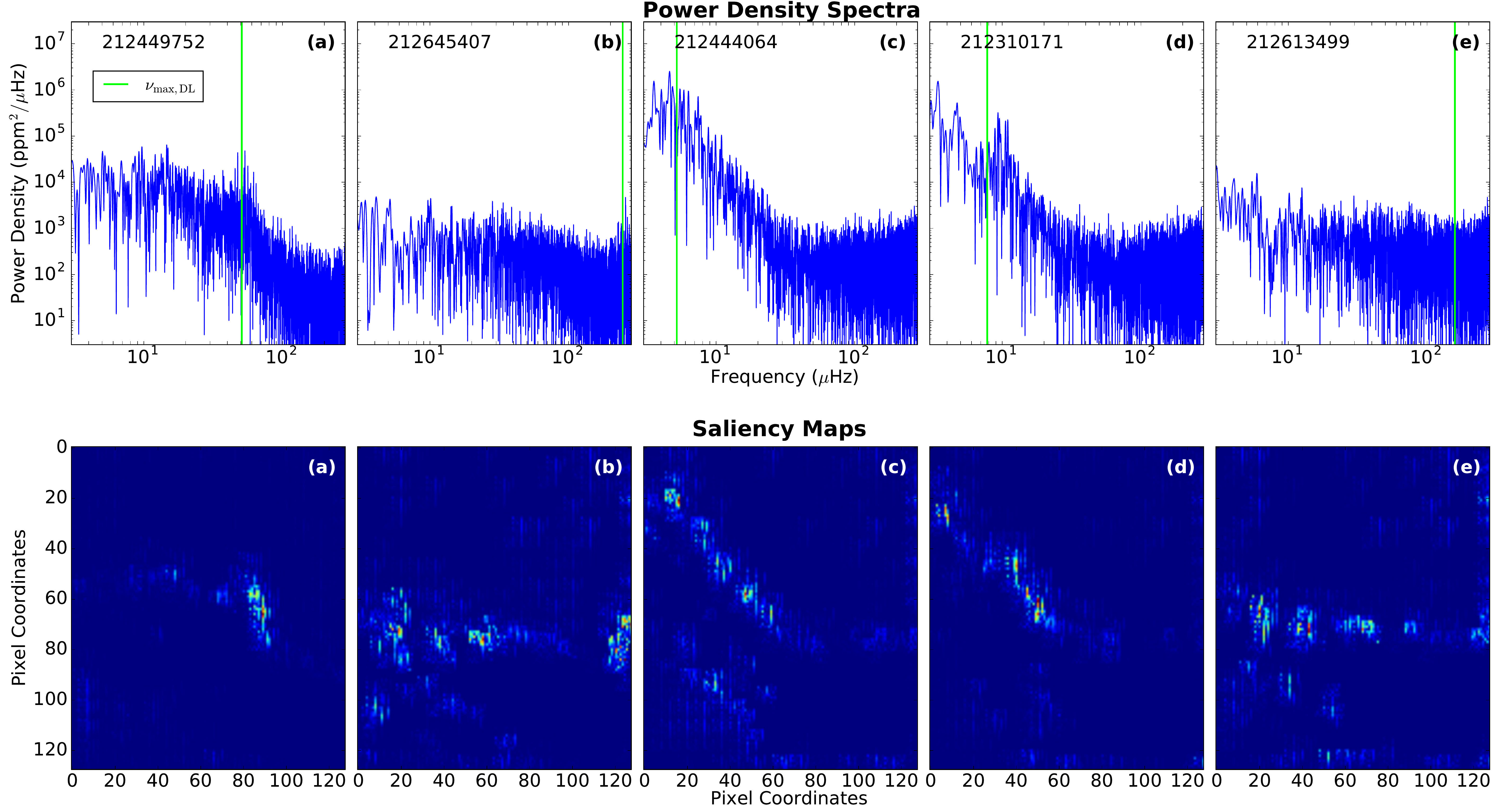}
	\caption{Power spectra of a subset of the outliers in Figure \ref{regoutliers} with their corresponding saliency maps by the regression model. `Hotter' regions in the saliency map highlight regions which contribute positively to the prediction of $\nu_{\mathrm{max, DL}}$. (a-c) The regression model predicts $\nu_{\mathrm{max}}$ correctly. (d-e) The regression model fails to predict $\nu_{\mathrm{max}}$ correctly.}
	\label{saliencyregression}
\end{figure*}
The middle row in Figure \ref{regoutliers} shows cases where $\nu_{\mathrm{max, BAM}}$ is the correct $\nu_{\mathrm{max}}$ estimate. In these cases, $\nu_{\mathrm{max, DL}}$ tends to be predicted slightly off-center from the center of the power excess envelope. It is possible that the regression model does in fact recognize the power excess, however its extracted features of that particular image are not sharply defined spatially, resulting in an inaccurate $\nu_{\mathrm{max}}$ estimate. Nonetheless, such cases are rare. The bottom row in Figure \ref{regoutliers} shows cases where neither $\nu_{\mathrm{max}}$ estimate is correct. In Figure \ref{regoutliers}k, we see $\nu_{\mathrm{max, DL}}$ and $\nu_{\mathrm{max, BAM}}$ offset from the true $\nu_{\mathrm{max}}$ between both estimates. Figures \ref{regoutliers}l-n show incorrect $\nu_{\mathrm{max}}$ estimates for stars with a true $\nu_{\mathrm{max}}$ close to the long-cadence Nyquist frequency, $\nu_{\mathrm{Nyq}}\simeq283\mu$Hz. While generally in the correct range of frequencies ($\nu_{\mathrm{max}} \apprge 200\mu$Hz), $\nu_{\mathrm{max, DL}}$ for these stars are not close enough to the true $\nu_{\mathrm{max}}$ to be good estimates. While the regression model is capable of correctly predicting a true $\nu_{\mathrm{max}}$ close to $\nu_{\mathrm{Nyq}}$ (see Figure \ref{regoutliers}a), we infer that its inaccuracy in Figures \ref{regoutliers}l-n is caused by the high-frequency power excess not being highly prominent in the image. Finally, Figure \ref{regoutliers}o shows a case where the true $\nu_{\mathrm{max}}$ cannot be determined visually from the images, because the power spectrum profile is atypical of a solar-like oscillator. Visual inspection of the power spectrum of this star in the linear scale reveals there is potentially a power excess with $\nu_{\mathrm{max}}\simeq19\mu$Hz. Nevertheless, neither $\nu_{\mathrm{max, DL}}$ nor $\nu_{\mathrm{max, BAM}}$ provided correct predictions.

\subsection{Regression Model Visualizations}  

We show the saliency maps of the regression model in Figure \ref{saliencyregression}, where the shown examples are subsets of the outliers in Figure \ref{regoutliers}. We see that while these maps are slightly noisy, they are very similar in appearance to that of the classifier model (Figures \ref{saliencyworst},  \ref{saliencydetection}, \ref{saliencynondetection}), in which highlighted features in the saliency map contribute towards the prediction of $\nu_{\mathrm{max, DL}}$. For correct predictions of $\nu_{\mathrm{max}}$ (Figures \ref{saliencyregression}a-c), the regression model appears to correctly place emphasis on the location of the power excess in the image. Additionally, other regions besides the power excess also appear to give contextual clues, such as the sharp granulation slope in Figure \ref{saliencyregression}c. In Figure \ref{saliencyregression}d, we see that although emphasis is correctly placed on the power excess, the $\nu_{\mathrm{max, DL}}$ is offset from its center. Finally, in Figure \ref{saliencyregression}e, we see that the power excess near $\nu_{\mathrm{Nyq}}$ is barely highlighted in the saliency map, in agreement with our argument that this $\nu_{\mathrm{max, DL}}$ inaccuracy is due to the lack of prominence of the power excess within the image.

\section{Conclusions}
With a focus on K2 data, we have developed a deep learning classifier model to perform efficient detection of red giants showing solar-like oscillations. This model learns and predicts on 2D images of the power spectra, hence does not require model fitting, making it very robust. By comparing the classifier's predictions to K2 targets that have been classified by expert visual inspection, the classifier achieved up to 98.1\% accuracy on the validation set comprising Campaign 6 stars, while it scored a 99.1\% accuracy on the test set comprising Campaign 3 stars. We presented visualizations of the classifier model and discovered that it observes the `correct' features for detecting solar-like oscillations, while it attributes sharp, vertical peaks and flat white noise to non-detections, which can potentially lead it to falsely classify very luminous red giants with $\nu_{\mathrm{max}} \apprle5\mu$Hz as non-detections.   

We also developed a deep learning regression model to provide an estimate of $\nu_{\mathrm{max}}$ for oscillating red giants, also using 2D images of the power spectra. We predicted on red giants in K2 Campaigns 3 and 6, where we estimated the uncertainty of the regression model, $\sigma_{\mathrm{DL}}$, to be about 5\%. We also compared our results to measured $\nu_{\mathrm{max}}$ values from the BAM pipeline and estimated the standard deviation of $\nu_{\mathrm{max, DL}}$ from $\nu_{\mathrm{max, BAM}}$ to be about 5-7\%. Additionally, for Campaigns 6 and 3, we compared $\nu_{\mathrm{max, DL}}$ to BAM's $\nu_{\mathrm{max}}$ uncertainties, $\sigma_{\mathrm{BAM}}$, and found that $\sim$20\% of our predictions lie within 1$\sigma_{\mathrm{BAM}}$, and $\sim$40\% lie within 2$\sigma_{\mathrm{BAM}}$, which is comparable to a human-level performance.

We found 229 out of 1747 $\nu_{\mathrm{max, DL}}$ values lie outside $\pm20\%$ of $\nu_{\mathrm{max, BAM}}$ for Campaign 6, and determined that $\nu_{\mathrm{max, DL}}$ is correct for 163 such outliers. We only found 9 such outliers out of 525 predictions for Campaign 3, where we determined that $\nu_{\mathrm{max, DL}}$ provided the right estimate for 6 of them. Hence, in most cases our regression model produces the correct $\nu_{\mathrm{max}}$ prediction, although it may potentially provide not very accurate estimates for low amplitude oscillations with $\nu_{\mathrm{max}}$ near $\nu_{\mathrm{Nyq}}$. Nonetheless, the combined application of the classifier and regression models present a method highly capable of increasing the asteroseismic yield from each K2 Campaign by complementing the analysis from existing asteroseismic pipelines. Our results also show great promise for applying similar techniques to the millions of stars observed by TESS starting in 2018.

\section*{Acknowledgements}
Funding for this Discovery mission is provided by NASA's Science Mission Directorate. We thank the entire \textit{Kepler} team without whom this investigation would not be possible. D.S. is the recipient of an Australian Research Council Future Fellowship (project number FT1400147). We would also like to thank Timothy Bedding, Daniel Huber, and the asteroseismology group at The University of Sydney for fruitful discussions.

\bibliography{bibi3}



\end{document}